\documentclass[aps,prc,twocolumn,superscriptaddress,showpacs,longbibliography,lengthcheck]{revtex4-1}
 
\usepackage{amsmath,amssymb}
\usepackage{graphicx}
\usepackage{color} 
\usepackage{txfonts}

\usepackage{ulem}

\usepackage{xspace}  
\usepackage{multirow}
\usepackage{dcolumn}

\begin{document}


\title{Emerging concepts in nuclear structure based on the shell model} 

\newcommand{\aut}{         \affiliation{Department of Physics, The University of Tokyo, 7-3-1 Hongo, Bunkyo, Tokyo 113-0033, Japan}}
\newcommand{\acns}{        \affiliation{Center for Nuclear Study, The University of Tokyo, 7-3-1 Hongo, Bunkyo, Tokyo 113-0033, Japan}}
\newcommand{\ariken}{      \affiliation{RIKEN Nishina Center, 2-1 Hirosawa, Wako, Saitama 351-0198, Japan}}
\newcommand{\ajar}{         \affiliation{Advanced Science Research Center, Japan Atomic Energy Agency, Tokai, Ibaraki 319-1195, Japan} }

 \newcommand{\aemt}{\email{Corresponding author: otsuka@phys.s.u-tokyo.ac.jp}}  
 
\author{Takaharu~Otsuka}   \aemt  \aut \ariken \ajar
\

\date{\today}

\begin{abstract}  
Some emerging concepts of nuclear structure are overviewed.  (1) Background: the many-body quantum structure of atomic nucleus, a complex system comprising protons and neutrons (called nucleons collectively), has been studied largely based on the idea of the quantum liquid ({\it a la} Landau), where nucleons are quasiparticles moving in a (mean) potential well, with weak ``residual'' interactions between nucleons.   The potential is rigid in general, although it can be anisotropic.  While this view was a good starting point, 
it is time to look into kaleidoscopic aspects of the nuclear structure brought in by underlying dynamics and nuclear forces.  (2) Methods: exotic features as well as classical issues are investigated from fresh viewpoints based on the shell model and nucleon-nucleon interactions.  The 70-year progress of the shell-model approach, including effective nucleon-nucleon interactions, enables us to do this.  (3) Results: we go beyond the picture of the solid potential well by activating the monopole interactions of the nuclear forces.  This produces notable consequences in key features such as the shell/magic structure, the  shape deformation, the dripline, {\it etc}.   These consequences are understood with emerging concepts such as shell evolution (incl. type-II), T-plot, self-organization (for collective bands), triaxial-shape dominance, new dripline mechanism, {\it etc}.  The resulting predictions and analyses agree with experiment.  (4) Conclusion: atomic nuclei are surprisingly richer objects than initially thought.
\end{abstract}




\maketitle

\section{Introduction \label{intro}}

The atomic nucleus is in a unique position in physics that it is an isolated object but comprises many quantum ingredients.    Some emerging concepts for the structure of atomic nuclei will be overviewed in this article, focusing on the works in which the author was involved.
Obviously, those concepts have been found or clarified thanks to the great progress of nuclear structure physics over 70 years, including the shell model.

In fact, the understanding of nuclear structure is based, to a great extent, on the shell model, which was introduced by Mayer \cite{mayer1949} and Jensen \cite{haxel1949} in 1949.  Since then, the shell model has been developed significantly in many ways: an initial phase as many-body physics was presented, for instance, by Talmi in \cite{talmi1962}, in contrast to Mayer-Jensen's independent-particle model.  The subsequent developments are reviewed, for instance by Caurier {\it et al.} in \cite{caurier2005} up to 2005, and in this volume up to date. I would like to sketch emerging concepts of nuclear structure based on recent shell-model studies involving the author, as many other studies are to be presented in other articles of the same volume.

The atomic nucleus comprises $Z$ protons and $N$ neutrons.  Their sum is called the mass number $A = Z + N$.  Among atomic nuclei, stable nuclei are characterized by their infinite or practically infinite life times, and are characterized by rather balanced $Z$ to $N$ ratios, with $N/Z$ ranging from about 1 up to about 1.5.  There are about 300 nuclear species of this category.  Other nuclei are called exotic (or unstable) nuclei.  The total number of them is unknown, but seems to be between 7000 and 10000, providing a huge show window of various features as well as the paths of nucleosynthesis in the cosmos (see, for instance,  \cite{gade2008,sorlin2008,nakamura2017}).  The exotic nuclei decay,  by $\beta$ ({\it i.e.} weak) processes, to other nuclei where $Z$ and $N$ are better balanced, as the $\beta$ decay alters a neutron to a proton or {\it vice versa}.  This decay occurs successively, until the process terminates at a stable nucleus.  Thus, only stable nuclei exist on earth, while exotic nuclei do not, being exotic literally. 

Some of the emerging concepts were conceived in the study of exotic nuclei, particularly by looking at the shell structure and magic numbers of them.  The obtained concepts were found later not to be limited to exotic nuclei.   In this way, after the initial trigger by exotic nuclei, the overall picture of the nuclear shell structure has been renewed, and Sec. 2 of this article is devoted to a sketch of it with two major keywords, the {\it monopole interaction} and the {\it shell evolution}.  

We then focus on the deformation of nuclear surface.  The surface deformation from the sphere has been very important subjects since 1950's, as initiated by Rainwater \cite{rainwater1950}, and by Bohr and Mottelson in \cite{bohr1952,bohr_mottelson1953,BMbook1,bohr_mottelson_book2}.
In particular, the shape coexistence phenomenon is discussed as the crossroad between the shell evolution and the deformation, leading to the concept of {\it type II shell evolution}.
Although I will not discuss extensively on the methodology of the shell model calculation in this article because of the length limitation, the {\it T-plot} of the Monte Carlo Shell Model will be mentioned as an essential theoretical tool for many physics cases of this article. These are main subjects of Sec. 3.

The in-depth clarification of the collective band is connected to the fundamental question on the relation between the single-particle degrees of freedom and the collective motion of nucleons.  These two must be connected through nuclear forces.   This question has not been clarified enough as also addressed by G. E. Brown \cite{schaefer2014}.  I shall focus, in Sec. \ref{coll}, how this question may be understood more deeply, by introducing the {\it self-organization} aspect of the collective bands and by raising the importance of the triaxiality of nuclear shapes including the ground states.

The interplay between the monopole interaction and the quadrupole deformation is shown to be a major mechanism of the determination of the neutron driplines.  This approach explains neutron driplines observed recently.  We are led to two dripline mechanisms: the traditional one with the single-particle origin and the present one.   The monopole-quadrupole interplay responsible for this new dripline mechanism is explained in detail in Sec. \ref{drip}.  As an alternative case, spherical isotopes, such as Ca, Ni, Sn and Pb, are predicted to have longer isotopic chains.

The intention of this article is to show the major flow of basic ideas and related results without going into details.  I hope that the reader can grasp this flow, and could become interested in watching further developments.  The past 70 years are really great for the shell model, but the coming years look equally or even more brilliant.  I apologize for not covering many of major developments in the last 70 years, as such coverage is not possible within this article but the other articles of this volume are expected to help.

\section{Shell Evolution due to Monopole Interaction \label{sec:evolution}}

\subsection{Mayer-Jensen's shell model and observed magic numbers \label{MJ}}
\begin{figure*}[bt]
\includegraphics[width=11 cm]{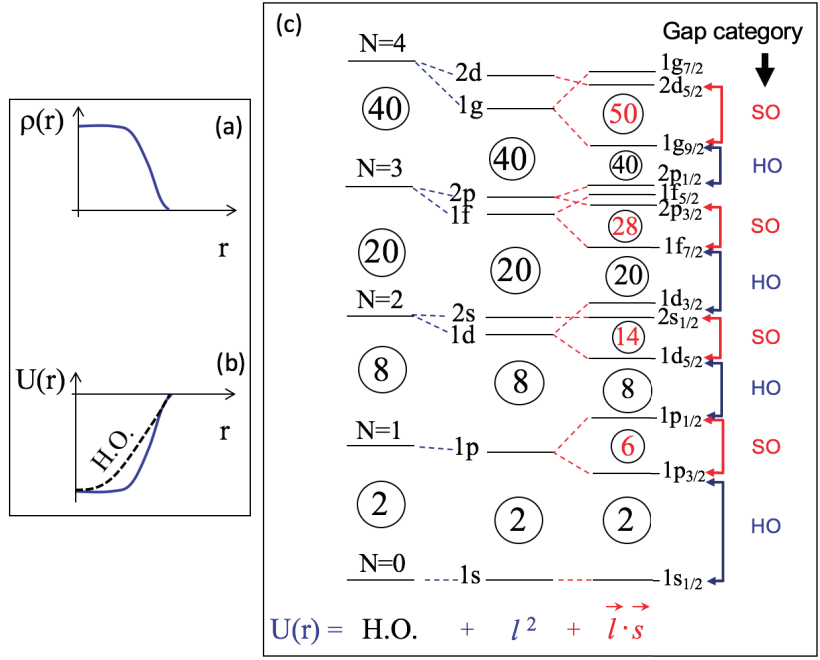}
\caption{Schematic illustration of (\textbf{a})  density distribution of nucleons in atomic nuclei, 
(\textbf{b})  a mean potential (solid line) produced by nucleons in atomic nuclei and an approximation by a Harmonic Oscillator (HO) potential (dashed line).  In (a) and (b), the horizontal line denotes the radius from the center of the nucleus.
(\textbf{c})  The shell structure produced with resulting magic numbers in circles.  (left column)  Only the HO potential is taken with HO quanta shown as N=0, N=1, 
...  (N here does not mean the neutron number, $N$.)
(middle column) The so-called $\ell ^2$ term is aded to the HO potential, where the magic gaps are shown in circles.  The single-particle orbits are labeled in the standard way to the left.   (right column) The spin-orbit term, $(\vec{l} \cdot \vec{s})$, is included further, and magic gaps emerging from this term are shown in red.    
The single-particle orbits are labeled to the right, including $\vec{j}=\vec{l}+\vec{s}$.  The magic gaps are classified as ``HO'' and ``SO" for the HO potential and spin-orbit origins, respectively.  
Taken from Fig. 2 of \cite{otsuka2020}, which was adopted from \cite{ragnarsson1995}. 
\label{fig1}}
\end{figure*}

Mayer \cite{mayer1949} and Jensen \cite{haxel1949} proposed, in 1949, the model of the shell structure and magic numbers of atomic nuclei.  This model provided major guides for deeper and wider understanding of the structure of atomic nuclei.  While this is a similar situation to electrons in atoms, there are some differences.   Figure~\ref{fig1} depicts the basic idea and consequences of the Mayer-Jensen's scheme.  We start with the nuclear matter composed of protons and neutrons.  This matter shows an almost constant density of nucleons (collective name of protons and neutrons) inside the surface which is a sphere as a natural assumption (see Fig.~\ref{fig1} ({\bf a})).  Because of the short-range character of nuclear forces, this constant density results in a mean potential with a constant depth inside the surface, as shown 
in Fig.~\ref{fig1} ({\bf b}).   Let's assume that the density distribution is isotropic, producing an isotropic mean potential.  Figure~\ref{fig1} ({\bf b}) also suggests that the Harmonic Oscillator (HO) potential is a good approximation to this mean potential as long as the mean potential shows negative values as a function of $r$, the radius from the center of the nucleus.  We then switch from the mean potential to the HO potential, which is analytically more tractable.
Thus, the HO potential can be introduced from the constant density (sometimes referred to as "density saturation") and the short-range attraction due to nuclear forces.  

The eigenstates of the HO potential are single-particle states shown in the far-left column of Fig.~\ref{fig1} ({\bf c}) with associated magic numbers and HO quanta, N.  These HO magic numbers do not change by adding the minor correction of the $\ell^2$ term, the scalar product of the orbital angular momentum $\vec{l}$  (see the second column from left in Fig.~\ref{fig1} ({\bf c}); for details see \cite{BMbook1}).  

The crucial factor introduced by Mayer and Jensen was the spin-orbit (SO) term, $(\vec{l} \cdot \vec{s})$, the effect of which is shown in the third column from left in Fig.~\ref{fig1} ({\bf c}).  
The two orbits with the same orbital angular momentum, $\ell$, and the same HO quanta are denoted as,
\begin{equation}
j_> = \ell + 1/2  \,\,\, \mathrm{and} \,\,\,  j_< = \ell - 1/2,
\label{eq:j>j<}
\end{equation}
where $1/2$ is due to the spin, $s$ = 1/2.  The notation of $j_>$ and $j_<$ will be used frequently in this article.
 The spin-orbit term, 
 \begin{equation}
v_{ls} \,=\, f (\vec{l} \cdot \vec{s}), 
\label{eq:ls}
\end{equation}
is added to the HO + $\ell^2$ potential, where $f$ is the strength parameter.  With $f <  0$ as is the case for nuclear forces, the $j_>$ state is lowered in energy, whereas the $j_<$ state  is raised.    
The value of $f$ is known empirically to be about -20$A^{-2/3}$ MeV (see eq. (2-132) of \cite{BMbook1}).  

The final pattern of the single-particle energies (SPE) is shown schematically in Fig.~\ref{fig1} ({\bf c}).
The single-particle states are labelled in the standard way up to their $j$ values, and 
both HO and spin-orbit magic gaps are indicated in black and red, respectively.   
The magic numbers have been considered to be $Z, N$ = 2, 8, 20, 28, 50, 82 and 126, because the effect of the spin-orbit term becomes stronger as $j$ becomes larger.  In fact, the magic numbers 28, 50, 82 and 126 are all due to this effect.  Instead, the HO magic numbers beyond 20 were considered to be absent or show only minor effects.  We shall look back on them, from modern views of the nuclear structure covering stable and exotic nuclei.  

\begin{figure}[bt]
\includegraphics[width=8.5 cm]{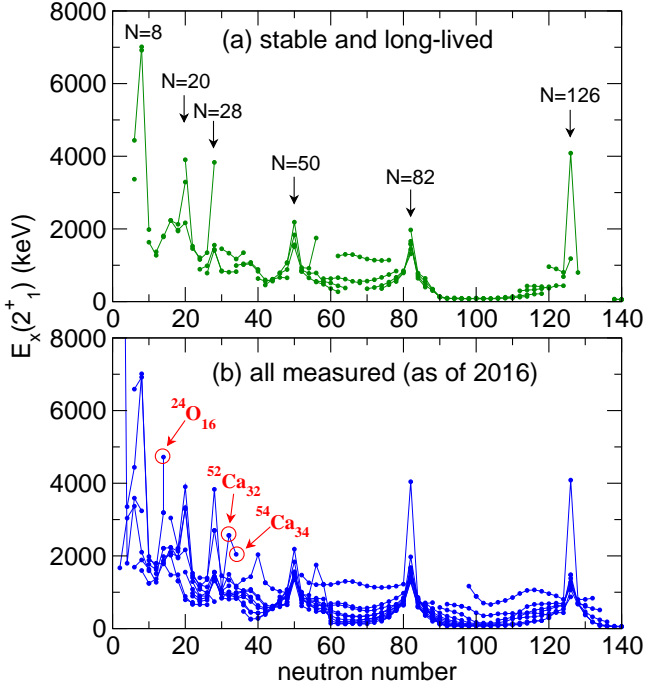}
\caption{ Systematics of the first 2$^+$ levels, for (\textbf{a})  stable and long-lived
nuclei, and  (\textbf{b}) all nuclei included measured up to 2016.  Taken from Fig. 4 of \cite{otsuka2020}.
\label{fig2}}
\end{figure}

We now investigate to what extent magic gaps in Fig.~\ref{fig1} ({\bf c}) have been observed.  
Figure~\ref{fig2} 
displays the observed excitation energies of the first 2$^+$ states of even-even nuclei as a function of $N$, where even-even stands for even-$Z$-even-$N$.   These excitation energies tend to be high at the magic numbers, because excitations across the relevant magic gap are needed.  The conventional magic numbers of Mayer and Jensen, $N$= 2, 8, 20, 28, ... 126 are expected to arise, and we indeed see sharp spikes at these magic numbers in Fig.~\ref{fig2}\textbf{a} where the excitation energies are shown for stable and long-lived ({\it i.e.} meta stable) nuclei.   
The panel \textbf{b} includes all measured first 2$^+$ excitation energies as of 2016.  In addition to the spikes in panel \textbf{a}, one sees some new ones.  One of them is at $N$=40, which corresponds to $^{68}$Ni$_{40}$, representing a HO magic gap at $N$=40.  There are three others corresponding to the nuclei, $^{24}$O$_{16}$, $^{52}$Ca$_{32}$ and $^{54}$Ca$_{34}$, as marked in red in the panel.   The 2$^+$ excitation energies of these nuclei are about a factor two higher than the overall trend, suggesting that $N$=16, 32 and 34 can be magic numbers, although none of them is present in Fig.~\ref{fig1} ({\bf c}).

These new possible magic numbers are consequences of what are missing in the argument for deriving magic gaps in Fig.~\ref{fig1} ({\bf c}).  We now turn to follow some passages along which this subject has been studied. 

\subsection{Monopole interaction}
The change from the Mayer-Jensen scheme is discussed from the viewpoint of the $NN$ interaction.  The Hamiltonian is written as,
\begin{equation}
\hat{H} \, = \, \hat{H}_0 + \, \hat{V} \, ,             
\label{eq:m_tot}
\end{equation}
where $\hat{H}_0$ denotes the one-body term given by
\begin{equation}
\hat{H}_0 \, =\, \Sigma_j \, \epsilon^{p}_{0;j} \, \hat{n}^p_j \, +\, \Sigma_j \, \epsilon^{n}_{0;j}  \, \hat{n}^n_j  \,,             
\label{eq:H_0}
\end{equation}
and $\hat{V}$ stands for the $NN$ interaction.  Here, $\hat{n}_j^{p,n}$ means the proton- or neutron-number operator for the orbit $j$, and $\epsilon^{p,n}_{0;j}$ implies 
proton or neutron single-particle energy (SPE) of the orbit $j$. This SPE is composed of the kinetic energy of the orbit $j$ and the binding energy on the orbit $j$ generated by all nucleons in the inert core.

\begin{figure*}[bt]
\includegraphics[width=16 cm]{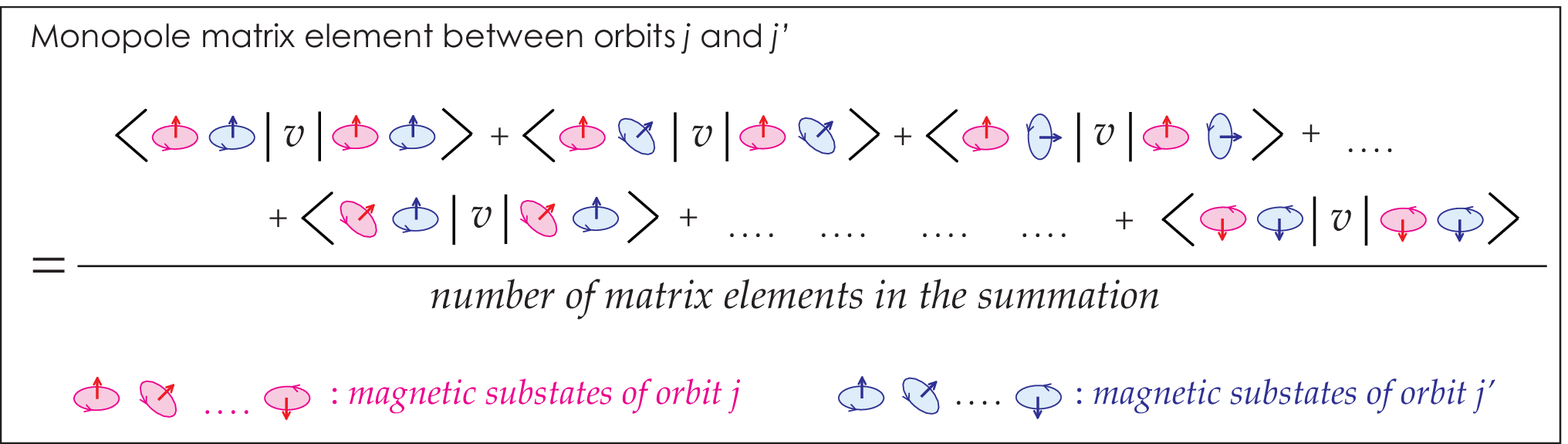}
\caption{Schematic illustration of the monopole matrix element.  Taken from Fig. 7 of \cite{otsuka2020}.
\label{mono}}
\end{figure*}

The interaction $\hat{V}$ can be decomposed, in general, into the two components: monopole and multipole interactions \cite{poves1981}, irrespectively of its origin, derivation or parameters.  The monopole interaction, denoted as $\hat{V}^{mono}$, is expressed in terms of the monopole matrix element, which is defined for single-particle orbits $j$ and $j'$ as,
\begin{equation}
\label{eq:mono}
V^{mono} (j,j') \, = \, \frac{\Sigma_{(m, m')}  \, \langle j,m\,;\, j',m' | \hat{V} | j,m \,;\, j',m' \rangle }{\Sigma_{(m,m')} \,1}, \\   
\end{equation}
where $m$ and $m'$ are magnetic substates of $j$ and $j'$, respectively, and the summation over $m, m'$ is taken for all ordered pairs allowed by the Pauli principle.  The monopole matrix element represents, as displayed schematically in Fig.~\ref{mono}, an orientation average for two nucleons in the orbits $j$ and $j'$.     See \cite{otsuka2020} for more detailed descriptions.

The monopole interaction between two neutrons is given as 
\begin{equation}
\label{eq:m_nn}
\hat{V}_{nn}^{mono}  \, = \,  \Sigma_{j} \, V_{nn}^{mono} (j,j) \, \frac{1}{2} \, \hat{n}_j^n \, (\hat{n}_j^n -1) + \Sigma_{j < j'} \, V_{nn}^{mono} (j,j') \, \hat{n}_j^n \, \hat{n}_{j'}^n  \, .
\end{equation}  
The monopole interaction between two protons is given similarly.
The monopole interaction between a proton and a neutron is given as 
\begin{equation}
 \hat{V}_{pn}^{mono} \, \approx \, \Sigma_{\, j , \,  j'}  \, \tilde{V}_{pn}^{mono} (j, \, j')\,  \,  \hat{n}^p_j \, \hat{n}^n_{j'}  \, , 
\label{eq:m_pn}
\end{equation}
where $\tilde{V}_{pn}^{mono} (j, \, j')$ stands for the monopole matrix element of the proton-neutron interaction slightly changed from the one given in eq.~(\ref{eq:mono}): changes are due to the isospin symmetry and are made only for $j$=$j'$  (see Sec. III A of \cite{otsuka2020} for details).    

The functional forms in eqs.~(\ref{eq:m_nn},\ref{eq:m_pn}) appear to be in accordance with the  intuition from the averaging over all orientations: no dependences on angular properties  ({\it e.g.} coupled $J$ values) between the two interacting nucleons, and the sole dependence on the number of particles in those orbits.

The (total) monopole interaction is written as
\begin{equation}
\hat{V}^{mono} \, = \, \hat{V}_{pp}^{mono}  + \hat{V}_{nn}^{mono} + \hat{V}_{pn}^{mono} \, ,            
\label{eq:Vmono}
\end{equation} 
and the monopole Hamiltonian is defined as,
\begin{equation}
\hat{H}^{mono} \, = \, \hat{H}_0 \, + \, \hat{V}^{mono} \, =\, \Sigma_j \, \epsilon^{p}_{0;j} \, \hat{n}^p_j \, +\, \Sigma_j \, \epsilon^{n}_{0;j}  \, \hat{n}^n_j \, + \, \hat{V}^{mono} \, .             
\label{eq:Hmono}
\end{equation}
The multipole interaction is introduced as
\begin{equation}
\label{eq:multi}
\hat{V}^{multi} \, = \,\hat{V} - \, \hat{V}^{mono} \, ,
\end{equation}
and the (total) Hamiltonian is written as  
$\hat{H}  \, =  \, \hat{H}^{mono} \, + \, \hat{V}^{multi}$.
The multipole interaction becomes crucial in many aspects of nuclear structure, for instance, the shape deformation, as touched upon in later sections of this article.   The monopole interaction has been studied over decades with many works, for example, \cite{bansal1964,baranger1970,poves1981,storm1983} (see \cite{otsuka2020} for more details).

We define the effective SPE (ESPE) of the proton (neutron) orbit $j$, denoted by $\hat{\epsilon}^{p}_{j}$ ($\hat{\epsilon}^{n}_{j}$), as the change of the monopole Hamiltonian, $\hat{H}^{mono}$ in eq.~(\ref{eq:Hmono}), due to the addition of one proton (neutron) into the orbit $j$.   This change is nothing but the difference, when $n^{p,n}_{j}$ is replaced by $n^{p,n}_{j}$+1.  For instance, the first term on the right-hand-side (RHS) of eq.~(\ref{eq:Hmono}) contributes to $\hat{\epsilon}^{p}_{j}$ by a constant, $\epsilon^{p}_{0;j}$.  As another example, the RHS of eq.~(\ref{eq:m_pn}) contributes 
by $\Sigma_{j'} \, \tilde{V}_{pn}^{mono} (j, j') \{(\hat{n}^p_j +1)\, \hat{n}^n_{j'}-\hat{n}^p_j \, \hat{n}^n_{j'} \} = \Sigma_{j'} \, \tilde{V}_{pn}^{mono} (j, j')  \hat{n}^n_{j'}$.   Combining all terms, the ESPE of the proton orbit $j$ is given as,  
\begin{equation}
\label{eq:ESPEproton}
\hat{\epsilon}^{p}_{j} \, = \, \epsilon^{p}_{0;j} \, + \,  \Sigma_{j'} \, V_{pp}^{mono} (j, \, j') \, \hat{n}^p_{j'} + \,  \Sigma_{j'} \, \tilde{V}_{pn}^{mono} (j, \, j') \, \hat{n}^n_{j'} \, .
\end{equation}
The second and third terms on the RHS are obviously contributions from valence protons and neutrons, respectively.  
The neutron ESPE is expressed similarly as
\begin{equation}
\label{eq:ESPEneutron}
\hat{\epsilon}^{n}_{j} \, = \, \epsilon^{n}_{0;j} \, + \, \Sigma_{j'} \, V_{nn}^{mono} (j, \, j') \, \hat{n}^n_{j'} + \,  \Sigma_{j'} \, \tilde{V}_{pn}^{mono} (j', \, j) \, \hat{n}^p_{j'}  \, .
\end{equation}
In many practical cases, an appropriate expectation value of the ESPE operator is also called the ESPE with an implicit reference to some state characterizing the structure, {\it e.g.} the ground state.  

The ESPE as an expectation value is often discussed in terms of the difference between two states, {\it e.g.} $\Psi$ and $\Psi'$. We here show the formulas for this difference.  First we introduce the symbol $\Delta {\mathcal O}$ for an operator $\hat{{\mathcal O}}$  implying the difference, $\langle \Psi \, | \hat{{\mathcal O}} | \,  \Psi \rangle - \langle \Psi' \, | \hat{{\mathcal O}} | \,  \Psi' \rangle$.  Such differences of the ESPE values are expressed as,
\begin{equation}
\label{eq:epj}
\Delta \epsilon^{p}_j  \,=  \Sigma_{j'} \, V_{pp}^{mono} (j,j') \, \Delta n^p_{j'}  \,
   + \Sigma_{j'}  \, \tilde{V}_{pn}^{mono} (j, \, j')\, \,  \Delta n^n_{j'}  \, , 
\end{equation}     
and 
\begin{equation}
\label{eq:enj}
\Delta \epsilon^{n}_j  \,= \Sigma_{j'} \, V_{nn}^{mono} (j,j') \,\Delta  n^n_{j'}  \, 
  + \Sigma_{j'} \, \tilde{V}_{pn}^{mono} (j', \, j)\, \Delta  n^p_{j'}  \, .  
\end{equation} 
If $\Psi'$ is a doubly closed shell and $\Psi$ is an eigenstate with some valence protons and neutrons on top of this closed shell, these quantities stand for the evolution of ESPE's as functions of $Z$ and $N$.  One can thus see various physics cases represented by $\Psi$ and $\Psi'$.
Such ESPE's can provide picturesque prospects and great helps in intuitive understanding without resorting to complicated numerical calculations.    The notion of the ESPE has been well utilized, for instance, in empirical studies in \cite{grawe2004,sorlin2008}, in certain ways related to the present article.
 
The interaction $\hat{V}$ can be decomposed into several parts according to some classifications.   
The discussions in this subsection can then be applied to each part separately: the monopole interaction of a particular part of $\hat{V}$ can be extracted, and its resulting ESPE's can be evaluated.  Examples are presented in the subsequent subsections.   

We note that the definition of the ESPE can have certain variants with similar consequences, for instance, the combination of $n^{p,n}_{j}-1/2$ and $n^{p,n}_{j}+1/2$ instead of $n^{p,n}_{j}$ and $n^{p,n}_{j}+1$.  Appendix \ref{A1} shows a note on the relation to Baranger's ESPE.


\subsection{Monopole interaction of the central force}
With these formulations, we can discuss a variety of subjects ranging from the shell structure, to the collective bands, and to the driplines.  Let us start with the shell structure.    
While the discussions in Sec.~\ref{MJ} are based on basic nuclear properties, some aspects are  missing.  One of them is the orbital dependences of the monopole matrix element.  This dependence generally appears, but shows up more crucially in certain cases. 
We first look into the monopole interaction of the central-force component of $NN$ interactions.   Because the $NN$ interaction keeps the character of short-range attraction after modifications or renormalizations, the monopole matrix elements gain large magnitudes with negative sign ({\it i.e.} attractive), if radial wave functions of the single-particle orbits, $j$ and $j'$ in eq.(~\ref{eq:mono}), are similar to each other.  This similarity is visible, if these orbits are spin-orbit partners ($j=j_>$ and $j'=j_<$) with the identical radial wave functions (see eq.~(\ref{eq:j>j<})), for instance $1f_{7/2}$ and $1f_{5/2}$.  Another example is the coupling between unique-parity orbits, such as $1g_{9/2}$ and $1h_{11/2}$, for which the radial wave functions are similar because of no radial node.
These cases were pointed out by Federman and Pittel \cite{FP1977}, as the total effect of the $^3S_1$ channel of the $NN$ interaction without the reference to the monopole interaction.

\subsection{Monopole interaction of the tensor force}
\begin{figure}[bt]
\includegraphics[width=8 cm]{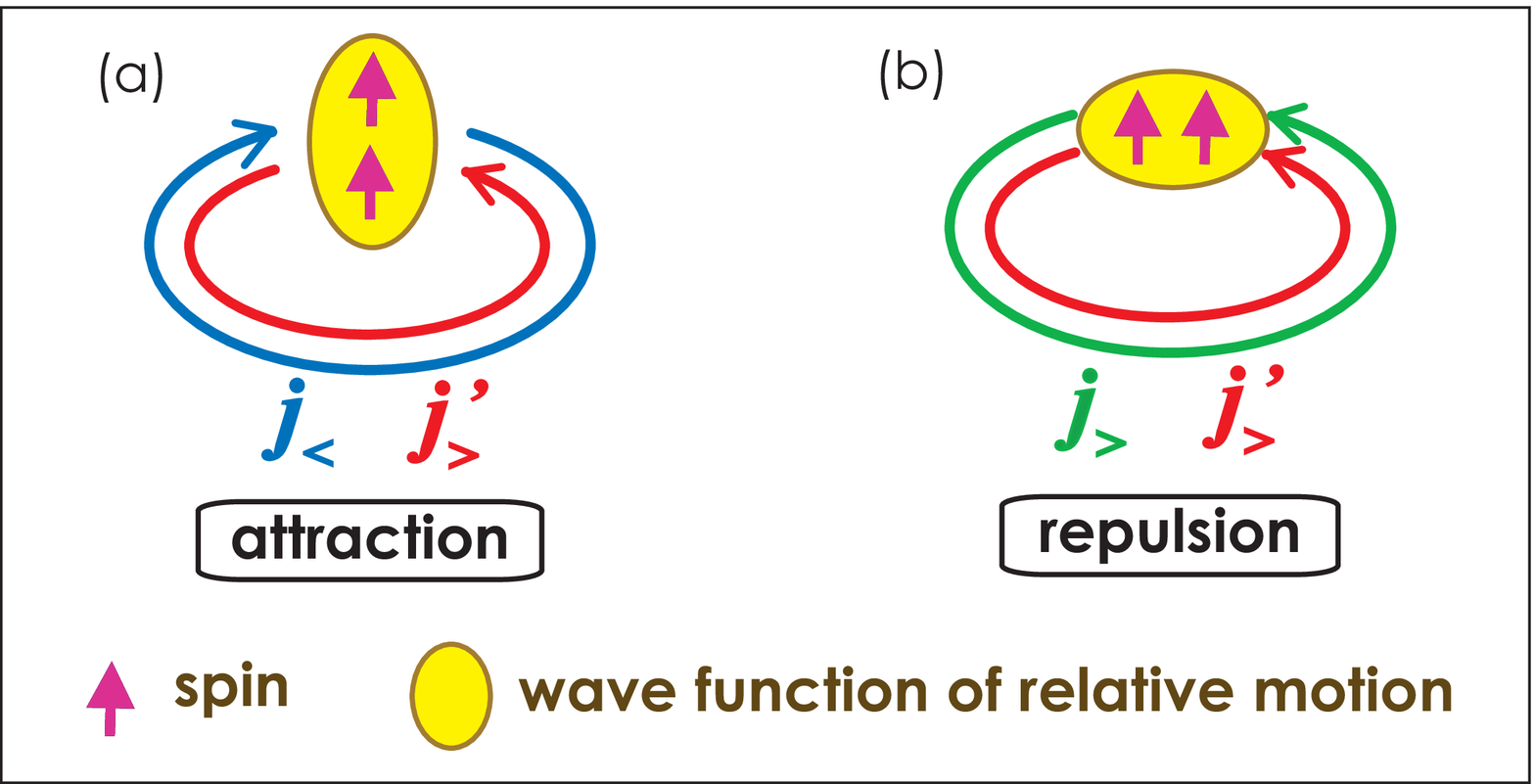}
\caption{Monopole interaction of the tensor force.  Taken from \cite{otsuka_2005}. }
\label{fig:ten1}
\end{figure}

Another important source of the monopole interaction with strong orbital dependences is
the tensor force.  The tensor force produces very unique effects on the ESPE.  This is shown in Fig.~\ref{fig:ten1}: the intuitive argument in \cite{otsuka_2005,otsuka2020} proves  that the monopole interaction of the tensor force is attractive between a nucleon in an orbit $j_<$ and another nucleon in an orbit  $j'_>$, whereas it becomes repulsive for combinations, ($j_>$, $j'_>$) or ($j_<$, $j'_<$).   The magnitude of such monopole interaction varies also.  For example, it is strong in magnitude between spin-orbit partners, or between unique-parity orbits, {\it etc.} \cite{otsuka2020}.  

The ESPE is shifted in very specific ways as exemplified in Fig.~\ref{fig:coexi} (b): if neutrons occupy a $j'_>$ orbit, the ESPE of the proton orbit $j_>$ is raised, whereas that of the proton orbit $j_<$ is lowered. 
This is nothing but a reduction of a proton spin-orbit splitting due to a specific neutron configuration.  
The amount of the shift is proportional to the number of neutrons in this configuration, as shown in eq.~(\ref{eq:epj}) and in Fig.~\ref{fig:coexi}(c).   
Other cases follow the same rule shown in Fig.~\ref{fig:ten1}.   
These general features have been pointed out in \cite{otsuka_2005} with an analytic formula and an intuitive description of its origin.

\subsection{Monopole-interaction effects from the central and tensor forces combined \label{subsec:vmu}}
 
The combined effects of the central and tensor forces were discussed in \cite{otsuka_2010} in terms of realistic shell-model interactions, USD \cite{usd} and GXPF1A \cite{gxpf1a}, referring to their roots in microscopic G-matrix $NN$ interactions proposed initially by Kuo and Brown \cite{kb1966,jensen_1995}.   Many other valuable shell-model interactions, for instance, (so-called) KB3 \cite{poves1981}, Kuo-Herling \cite{khint}, sn100pn \cite{sn100pn} and LNPS \cite{lnps} interactions, have been constructed from the G-matrix interactions sometimes with refinements like monopole adjustments.  The M3Y interaction \cite{m3y} is related to the G-matrix, too.  We appreciate the original contribution of the so-called G-matrix approach to the effective $NN$ interaction \cite{kb1966,jensen_1995}. 
 
\begin{figure*}[bt]
\includegraphics[width=15 cm]{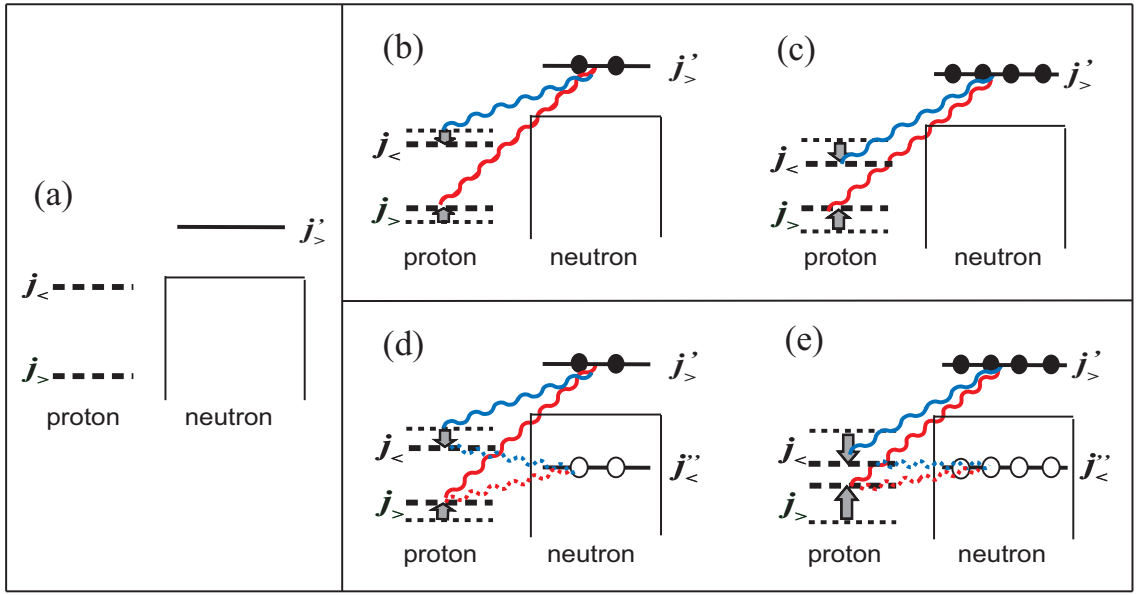}
\caption{Schematic picture of the ESPE change ({\it i.e.} shell evolution) due to the monopole interaction of the tensor force.  (a) SPE's with no neutrons in in the orbit $j'_>$.  (b) The shifts of the proton ESPE's due to two (valence) neutrons in the orbit $j'_>$.  (c) Same as (b) except for four neutrons.  (d,e)  Type-II shell evolution due to neutron particle-hole excitations.  Taken from \cite{otsuka_shapecoexi}.}
\label{fig:coexi}
\end{figure*}   

The {\bf V$_{\mathrm M \mathrm U}$ interaction} was then introduced as a general and simple shell-model $NN$ interaction. Its central part consists of Gaussian interactions with spin/isospin dependences, and their strength parameters are determined so as to simulate the overall features of the monopole matrix elements of the central part of USD \cite{usd} and GXPF1A \cite{gxpf1a} interactions.   Its tensor part is taken from the standard $\pi$- and $\rho$-meson exchange potentials \cite{otsuka_2005,tensor1,tensor2}.  Thus, the V$_{\mathrm M \mathrm U}$ interaction is defined as a function of the relative distance of two nucleons with spin/isospin dependences, which enables us to use it in a variety of regions of the nuclear chart, as we shall see.    A wide model space, typically a HO shell or more, is required in order to obtain reasonable results, though.

\begin{figure*}[bt]
\includegraphics[width=15.0 cm]{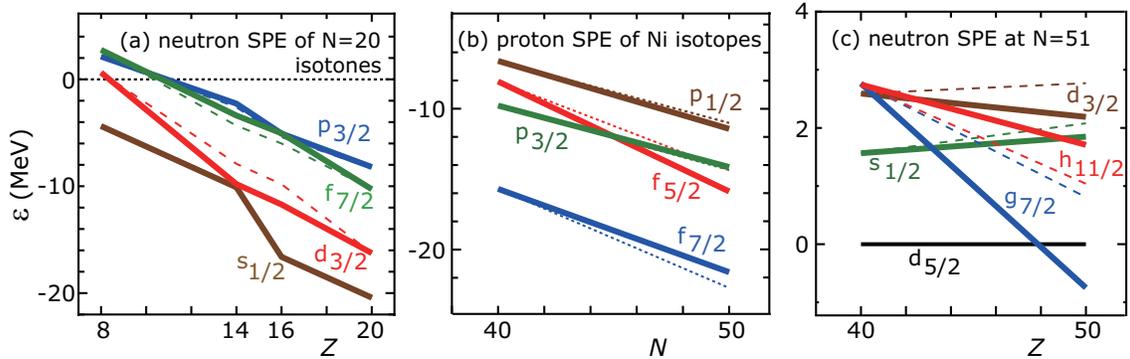}
\caption{Effective single-particle energies calculated by the VMU interaction. The dashed lines are obtained by the central force only, while the solid lines include both the central-force and the tensor-force contributions. 
Taken from \cite{otsuka_2010}. }
\label{fig:VMU}
\end{figure*}   

Figure~\ref{fig:VMU} depicts some examples: panel (a) displays the transition from a standard ({\it a la} Mayer-Jensen) $N$=20 magic gap to an exotic $N$=16 magic gap by plotting $\langle \hat{\epsilon}^{n}_{j} \rangle$ within the filling scheme (see eq.~(\ref{eq:ESPEneutron})), as $Z$ decreases from 20 to 8.  The tensor monopole interaction between the proton $d_{5/2}$ and the neutron $d_{3/2}$ orbits plays an important role.  
The small $N$=20 magic gap for $Z$=8-12 is consistent with the island of inversion picture (see reviews, {\it e.g.} \cite{caurier2005,otsuka2020}).
Panel (b) depicts the inversion between the proton $f_{5/2}$ and $p_{3/2}$ orbits as $N$ increases in Ni isotopes, by showing $\langle \hat{\epsilon}^{p}_{j} \rangle$ (see eq.~(\ref{eq:ESPEproton})).  The figure exhibits exotically ordered single-particle orbits for $N > 44$.  The tensor monopole interactions between the proton $f_{7/2,5/2}$ and the neutron $g_{9/2}$ orbits produce crucial effects.  Panel (c) shows significant changes of the neutron single-particle levels from $^{90}$Zr to $^{100}$Sn, in terms of $\langle \hat{\epsilon}^{n}_{j} \rangle$.   Without the tensor force, the degeneracy of $g_{7/2}$ and $d_{5/2}$ orbits in $^{100}$Sn does not show up.  


These changes of the shell structure as a function of $Z$ and/or $N$ were collectively called {\bf shell evolution} in \cite{otsuka_2005}.  The splitting between proton $g_{7/2}$ and $h_{11/2}$ in Sb isotopes shows a substantial widening as $N$ increases from 64 to 82 as pointed out by Schiffer {\it et al.} \cite{schiffer_2004}, which was one of the first experimental supports to the shell evolution partly because this was not explained otherwise.  Note that while the origin of the shell evolution can be any part of the $NN$ interaction, its appearance is exemplified graphically in Fig.~\ref{fig:coexi}(a,b,c) for the tensor force.   The shell-evolution trend depicted in Fig.~\ref{fig:VMU} appear to be consistent with experiment \cite{ensdf,otsuka_2010,sahin_2017,ichikawa_2019,liddick2006,seweryniak2007,otsuka2020}.  The monopole properties discussed in this subsection are consistent with the results shown in \cite{smirnova2010} obtained by the spin-tensor decomposition \cite{otsuka2020}.

\subsection{$N$=34 new magic number as a consequence of the shell evolution}


\begin{figure*}[bt]
\includegraphics[width=8.25 cm]{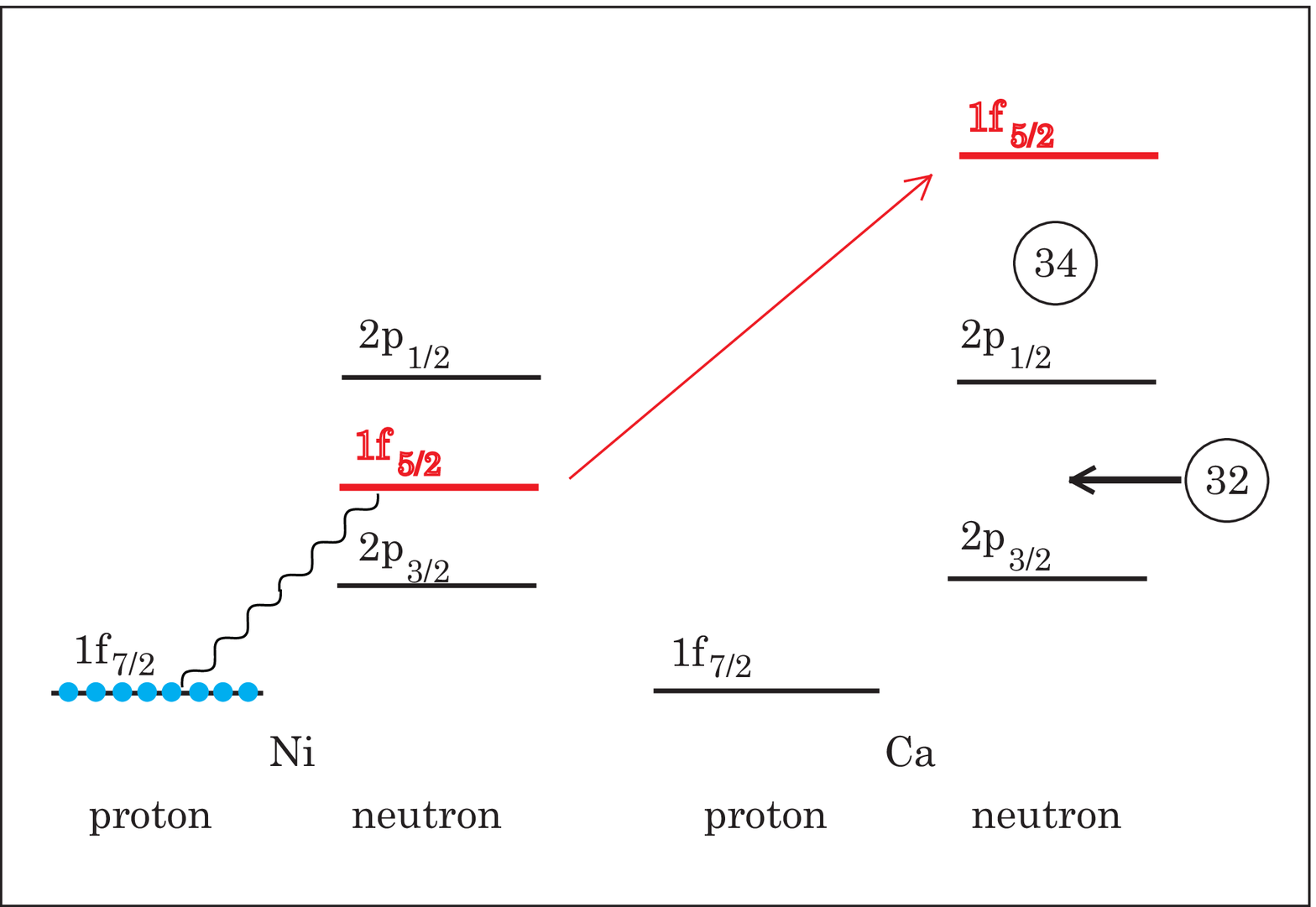}
\hspace{0.5cm}
\includegraphics[width=5.5 cm]{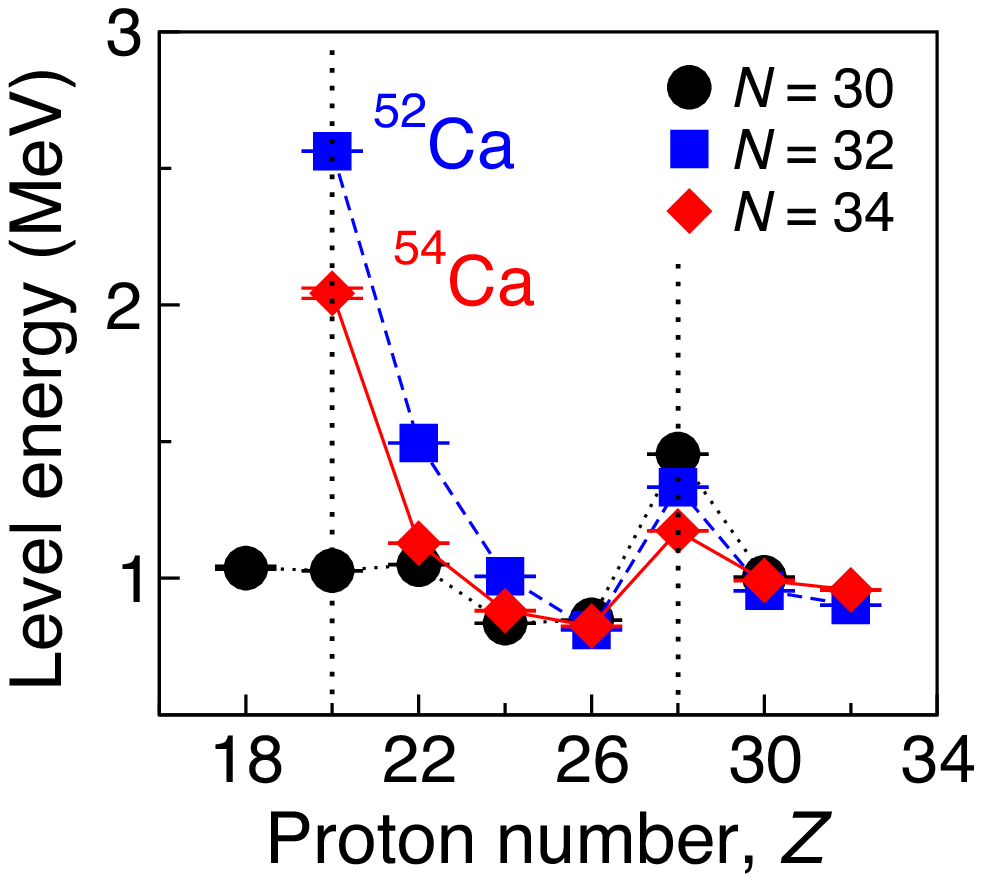}
\caption{(left) Schematic illustration of the shell evolution from Ni back to Ca for neutron
orbits. Blue circles denote protons. The wavy line is the interaction between the
proton 1$f_{7/2}$ orbit and the neutron 1$f_{5/2}$ orbit. The numbers in circles indicate 
magic numbers.  (right) Observed excitation energies of the 2$^+_1$ states.  Taken from Fig. 3 of \cite{otsuka_shapecoexi} and from Fig. 2 {\bf c} of \cite{steppenbeck_2013}. }
\label{fig:Ca}
\end{figure*}   

Among various cases of shell evolution, a notable impact was made by predicting a new magic number $N$=34.
Figure~\ref{fig:Ca} displays the shell evolution of some neutron orbits from Ni back to Ca isotopes, 
as $Z$ decreases from 28 to 20.   The 1$f_{5/2}$ orbit is between the 2$p_{3/2}$ and 2$p_{1/2}$
in Mayer-Jensen's shell model (see Fig.~\ref{fig1}).  By loosing eight protons in Ni isotopes, this canonical shell structure is destroyed as the 1$f_{5/2}$ orbit moves up above the 2$p_{1/2}$ orbit.
This movement of 1$f_{5/2}$ orbit creates the $N$=32 gap as a byproduct.  The energy shift of the 1$f_{5/2}$ orbit is due to the central and tensor forces by almost equal amounts: the $N$=34 magic gap does not appear without this shift, if the Mayer-Jensen scheme holds in Ni isotopes.   
The appearance of the $N$=34 magic number was predicted in \cite{otsuka_2001}, but 12 years were required \cite{janssens_2005} until the experimental verification became possible \cite{steppenbeck_2013} (see Fig.~\ref{fig:Ca} (right)).  The measured 2$^+$ energy levels are included in Fig.~\ref{fig2}(b).  More details are presented in \cite{otsuka2020}. Further evidences were obtained recently in \cite{michimasa2018,chen2019}.

\subsection{Monopole interaction of the 2-body spin-orbit force}

It is a natural question what effect can be expected from the 2-body spin-orbit force of the $NN$ interaction.  This force can be well described by the M3Y interaction, and its monopole effects were described in detail in \cite{otsuka2020}, particularly in its supplementary document.  Although the  monopole effects of this force contributes to the spin-orbit splitting \cite{otsuka2020}, the effect is 
much weaker than the tensor force in most cases, as also discussed in the article by Utsuno in this volume.   


\begin{figure}[bt]
\includegraphics[width=8.5 cm]{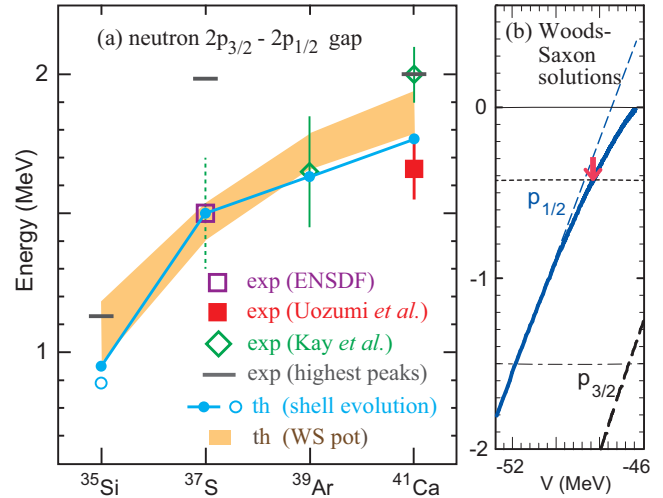}
\caption{(a) Neutron 2p$_{3/2}$-2p$_{1/2}$ splitting for
N = 21 isotones. The symbols are the centroids for
$^{37}$S \cite{ensdf}, $^{39}$Ar \cite{kay2017} and $^{41}$Ca \cite{uozumi1994} and \cite{kay2017}. The horizontal bars
are the energy differences between relevant highest peaks
\cite{burgunder2014}, \cite{kay2017}. Shell evolution predictions are shown by blue closed
symbols and solid line connecting them. The loose binding effect
for $^{35}$Si is included in the open circle. The calculation
with Woods-Saxon potential with parameters adjusted are
shown by the yellowish shaded area \cite{kay2017}. (b) Neutron 2p$_{1/2}$
single-particle energy (blue solid line) by a 
Woods-Saxon potential \cite{BMbook1} for varying depth parameter, V. The linear dependence of the deeply bound region is
linearly extrapolated (blue dashed line) and is compared to
the curved dependence that results from the proximity of the
continuum. The dashed line is for the 2p$_{3/2}$ orbit, and the loose-binding contribution to the present splitting appears to be 0.06 MeV against 1.5 MeV splitting itself. 
Taken from Fig. 8 of Supplementary Material of \cite{otsuka2020}. }
\label{fig:SiS}
\end{figure}   

An interesting case is found in the coupling between an $s$ orbits and the $p_{3/2,1/2}$ orbits.   There is no monopole effect from the tensor force, if an $s$ orbit is involved.
Instead, the $s$-$p$ coupling due to 2-body spin-orbit force can be exceptionally strong as intuitively stressed in \cite{otsuka2020}.  Figure~\ref{fig:SiS} shows that the possible significant change of the neutron 2$p_{3/2}$-2$p_{1/2}$ gap between $^{35}$Si and $^{37}$S is explained to a good extent by the shell evolution due to the 2-body spin-orbit force. 

\subsection{Monopole interaction from the three-nucleon force}


\begin{figure*}[bt]
\center
\includegraphics[width=9 cm]{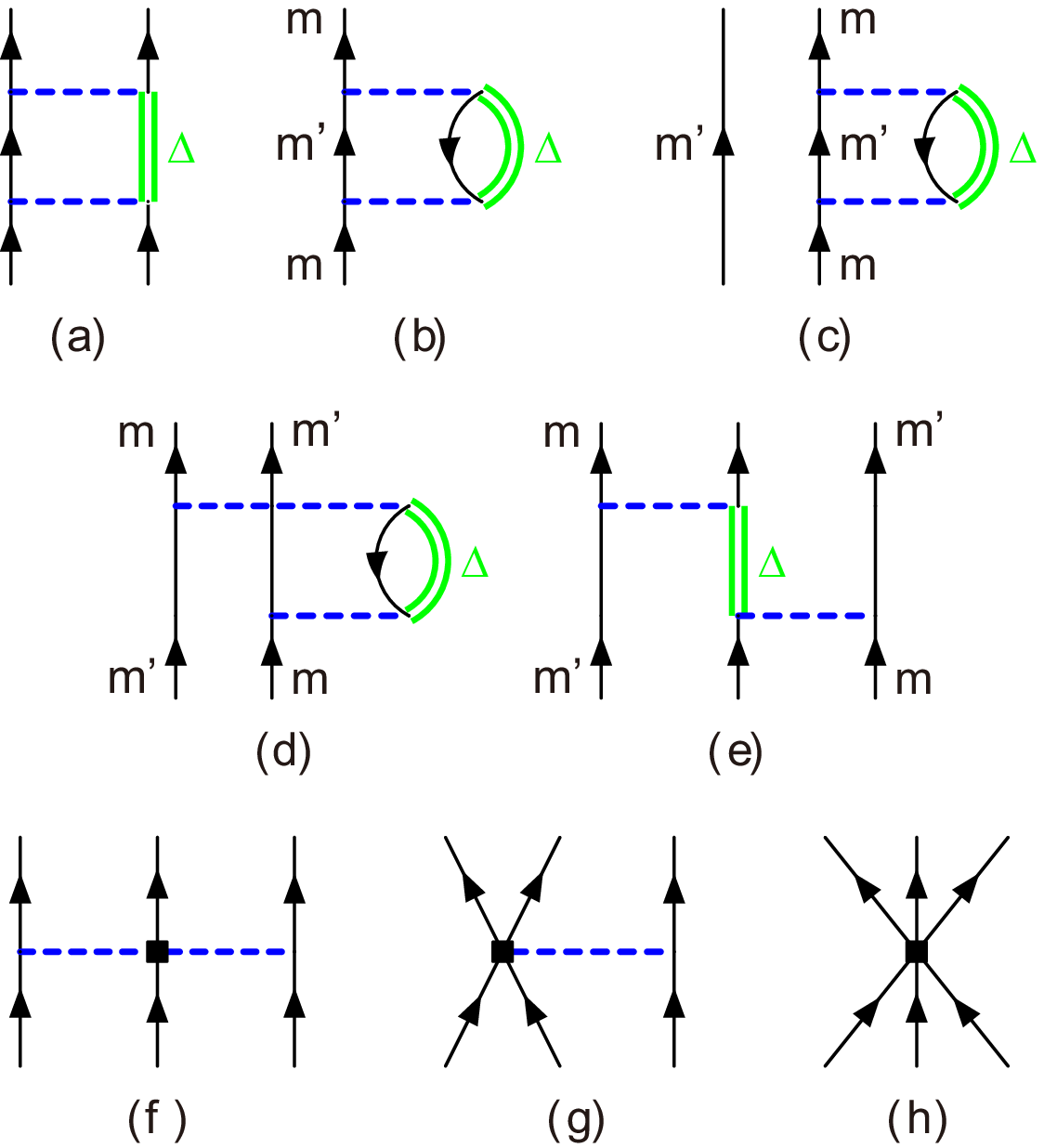}
\includegraphics[width=5 cm]{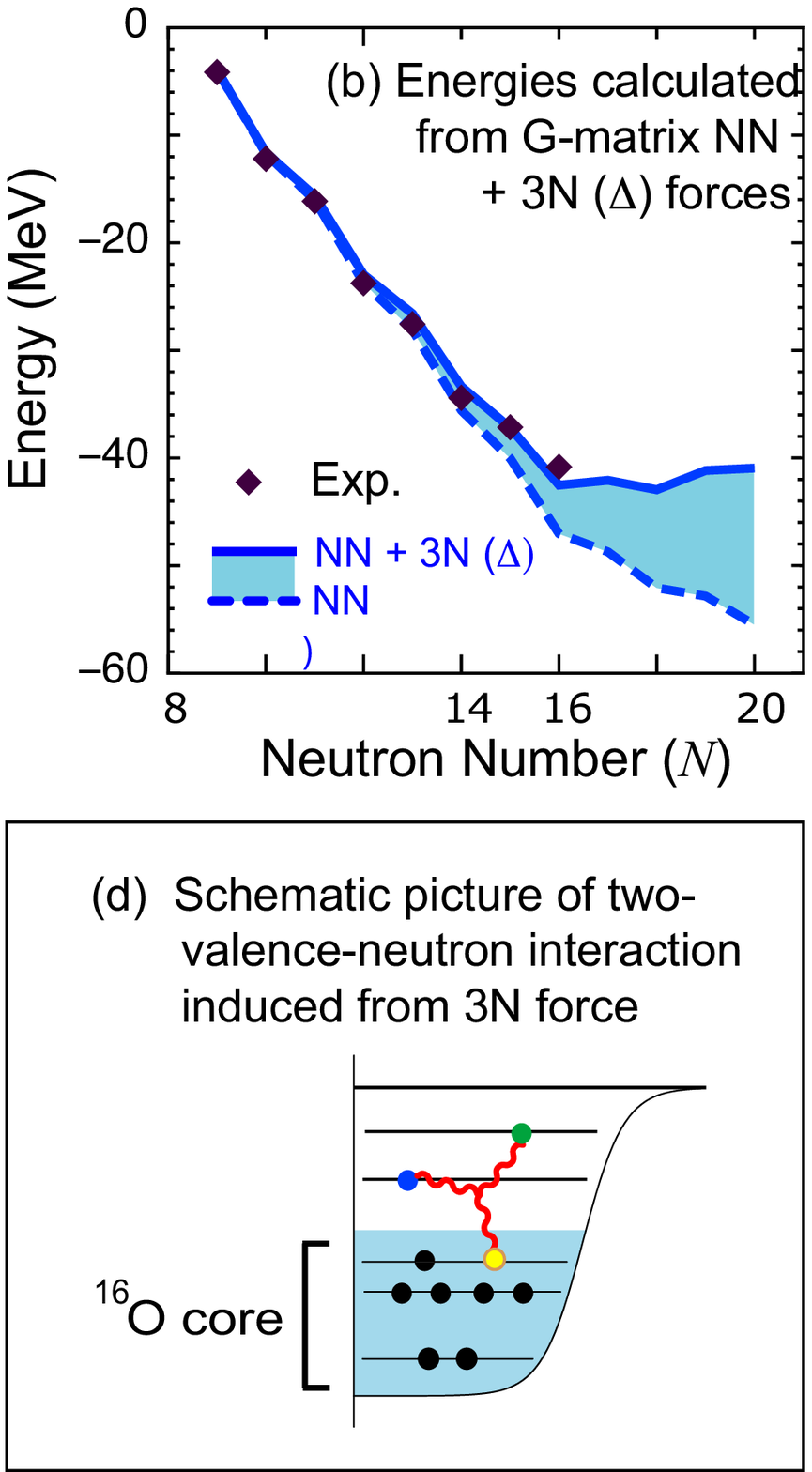}
\caption{Schematic illustration of the 3NF.  Based on Figs. 3 and 4 of \cite{otsuka_2010b}. }
\label{fig:3NF}
\end{figure*}   


The three-nucleon force (3NF) is currently of intense interest.  Among various aspects, we showed \cite{otsuka_2010b} the characteristic feature of the monopole interaction of the effective $NN$ interaction derived from the Fujita-Miyazawa 3NF.  Figure~\ref{fig:3NF} (a) displays the effect of the $\Delta$ excitation in nucleon-nucleon interaction.  The $\Delta$-hole excitation from the inert core changes the SPE of the orbit $j$ as shown in panel (b) where $m$ is one of the magnetic substates of the orbit $j$ and $m'$ means any state.  This diagram renormalizes the SPE, and observed SPE should include this contribution.  If there is a valence nucleon in the state $m'$ as in panel (c), the process in panel (b) is Pauli-forbidden.  However, in the shell-model and other nuclear-structure calculations, the SPE containing the effect of panel (b) is used.  One has to somehow incorporate the Pauli effect of panel (c), and a solution is the introduction of the process in panel (d).  In this process, the state $m'$ doubly appears in the intermediate state, but one can evaluate the Pauli effect by including panels (b) and (d) consistently.  This is a usual mathematical trick, and enables us to correctly treat the Pauli principle within the simple framework.  Panel (d) is equivalent to panel (e) which is nothing but the Fujita-Miyazawa 3NF, where the state $m'$ appears in double.  Similar treatment is carrried out in the chiral EFT framework.  Panel (f) corresponds to panel (e), but the violation of Pauli principle is a kind of hidden, because of a vertex in the middle (depicted by a square) instead of the $\Delta$ excitation.  

In this argument, the 3NF produces a repulsive monopole $NN$ interaction in the valence space, after the summation over the hole states of the inert core (see the right lower panel, labelled (d)).  The right upper panel indicates an example of the repulsive effect on the ground-state energy of oxygen isotopes, locating the oxygen dripline at the right place or solving the oxygen anomaly \cite{otsuka_2010b}.  This is rather strong repulsive monopole interaction, which is a consequence of inert core.  This means that the present case is irrelevant to the no-core shell model or other many-body approaches without the inert core ({\it e.g.} GFMC) \cite{QMC_RMP}.  This feature has caused some confusions in the past, but the difference is clear.   This repulsive effect is much stronger than the other effects of the 3NF \cite{dripline2020}.  I note that this repulsive effect was stressed by Talmi in 1960's \cite{talmi1962}.

\subsection{Short summary of Sec. \ref{sec:evolution}}

The shell evolution phenomena are seen in many isotopic and isotonic chains, and sometimes result in the formation of new magic gaps or the vanishing of old ones.  Figure~\ref{fig2}(b) displays the emergence of such new magic numbers $N$=16, 32 and 34, whereas the lowering of some 2$^+$ levels can mean the weakening of some magic numbers.  More changes may appear in the future studies.
Thus, the {\bf characteristic monopole features of the central, tensor, 2body-$LS$ and 3NF-based $NN$ interactions} and the {\bf resulting shell evolution} are among the {\bf emerging concepts of the nuclear structure}.   Interestingly, these findings are neither isolated nor limited to particular aspects, but are related to other aspects of the nuclear structure.  We now move on to such a case.

\section{Type-II Shell Evolution and Shape Coexistence \label{typeII}}

\subsection{Type-II Shell Evolution}


\begin{figure*}[bt]
\includegraphics[width=11 cm]{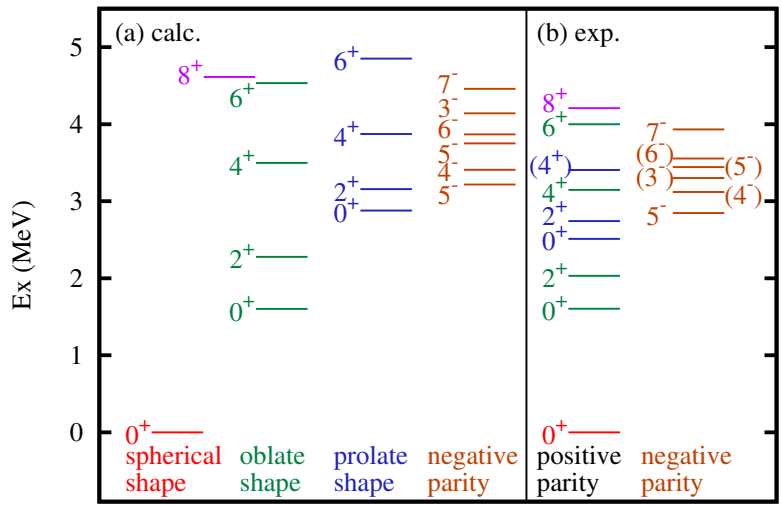}
\caption{Level scheme of $^{68}$Ni.  Taken from Fig. 2 of \cite{tsunoda2014}. }
\label{fig:Ni}
\end{figure*}   

The shell evolution shown in Fig.~\ref{fig:coexi}(b) and (c) are, respectively, due to the addition of two or four neutrons into the orbit $j'_>$.  Instead of adding, one can put neutrons into the orbit $j'_>$ by taking the neutrons from some orbits below $j'_>$, or equivalently by creating holes there, as shown in panel (d).  If such lower orbit happens to be the $j''_<$ orbit as in panel (d), its monopole matrix elements show just the opposite trends compared the $j'_>$ orbit.  However, because holes are created in $j''_<$, the sign of the monopole-interaction effect is reversed, and the final effect has the same sign as the monopole effect form the orbit $j'_>$ (see panel (d)).   Thus, the particle-hole (ph) excitation of the two neutrons in panel (d) reduces the proton $j_>$-$j_<$ splitting even more than in panel (b).  This reduction becomes stronger with the ph excitations of four neutrons, as depicted in panel (d).  Such strong reduction of the spin-orbit splitting produces interesting consequences beyond shell-structure changes. This type of the shell structure change within the same nucleus is called {\bf Type II Shell Evolution}.

\subsection{A doubly-closed nucleus $^{68}$Ni}
The Type-II shell evolution was first discussed in \cite{tsunoda2014} for $^{68}$Ni as an example.  Figure~\ref{fig:Ni} shows its theoretical and experimental energy levels. The theoretical results were obtained for the A3DA-m interaction by the {\bf Monte Carlo Shell Model}, which is a powerful methodology for the shell model calculation but is not discussed here due to the length limitation \cite{mcsm1,mcsm2,mcsm3,mcsm4}.  

Because $Z$=28 and $N$=40 of $^{68}$Ni are both magic (see Fig.~\ref{fig1}), the ground state is primarily a doubly closed shell.  Indeed, in the theoretical ground state, the occupation of the neutron $g_{9/2}$ orbit is negligibly small.  In contrast, the 0$^+_3$ state located at the excitation energy, Ex$\sim$ 3 MeV, is the band head of a rotational band of an ellipsoidal shape, and its neutron $g_{9/2}$ occupation number is as large as $\sim$4.  The mechanism shown in Fig.~\ref{fig:coexi}(e) is then switched on, reducing the proton $f_{5/2}$-$f_{7/2}$ splitting.  A reduced splitting facilitates more configuration mixing between these two orbits, which can produce notable effects on the quadrupole deformation as stated below.  
  
\subsection{Coexistence between spherical and deformed shapes \label{coexist}}

We here quickly overview the quadrupole deformation, or the shape deformation from a sphere to an ellipsoid \cite{bohr_mottelson_book2}.  
The quadrupole deformation is driven by the quadrupole interaction, a part of the multipole interaction in eq.~(\ref{eq:multi}).  The quadrupole interaction is a somewhat vague idea because of a certain mathematical complication, but its main effects can be simulated by the (scalar) coupling of the quadrupole moment operators.  
If the quadrupole moments are larger, {\it i.e.}, a stronger quadrupole deformation occurs, the nucleus gains more binding energy from the quadrupole interaction.   This is a very general phenomena, and because of this the ground states of many nuclei are deformed, although $^{68}$Ni is not among them.  

\begin{figure*}[bt]
\includegraphics[width=6.8 cm]{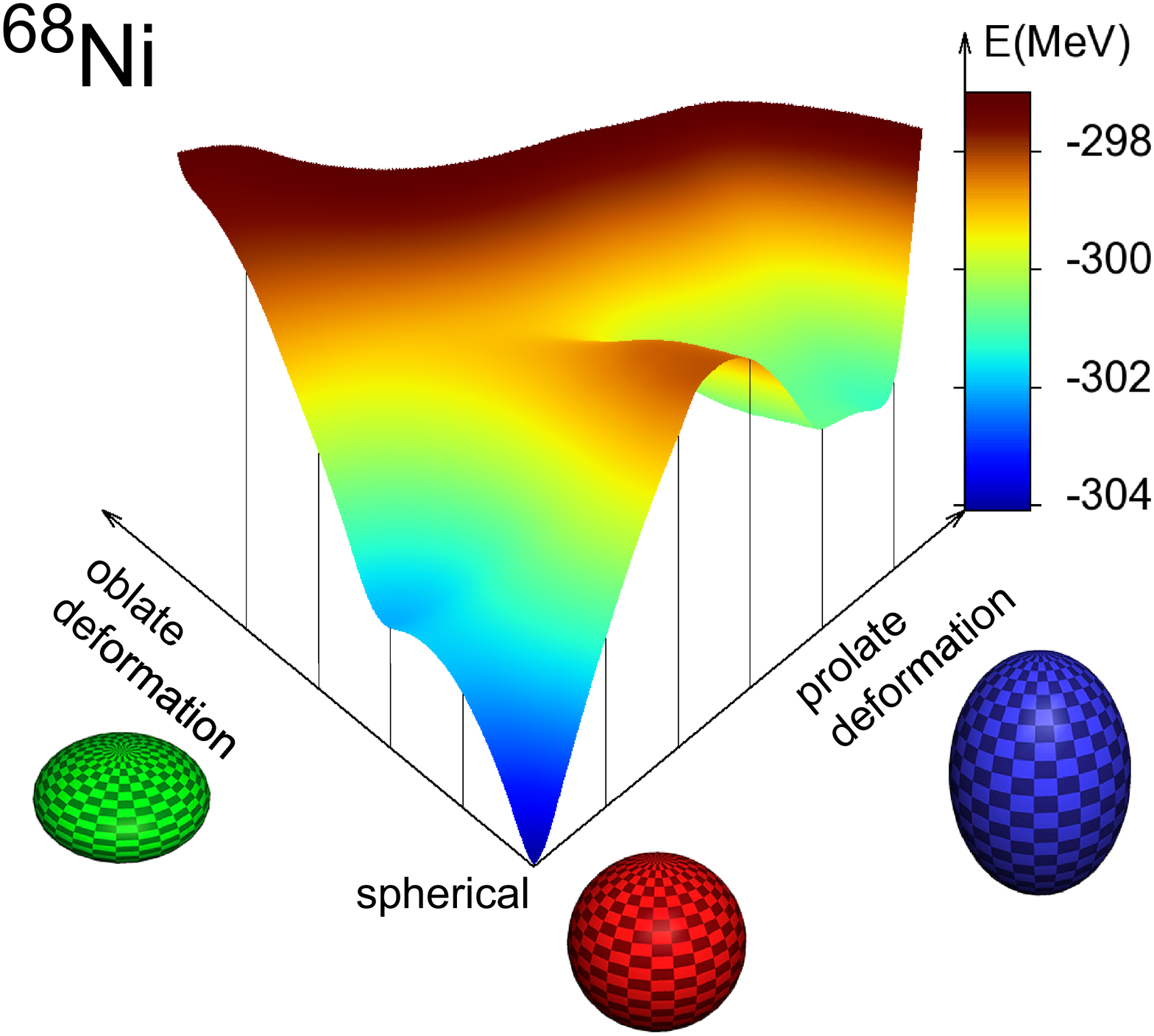}
\includegraphics[width=7.9 cm]{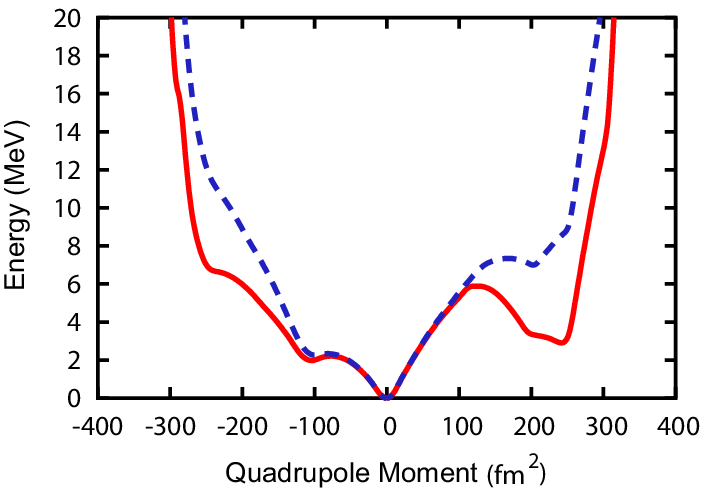}
\caption{(left) Potential energy surface (PES) of $^{68}$Ni.  Taken from Fig. 5 of \cite{otsuka2016}. 
(right) PES of $^{68}$Ni with axially symmetric shapes.  The solid line shows the PES of the full Hamiltonian, whereas the dashed line is the PES with practically no tensor-force contribution.  Taken from Fig. 6 of \cite{otsuka2016}.
}
\label{fig:Ni_pes}
\end{figure*}

The energy of $^{68}$Ni (intrinsic state) is graphically illustrated in Fig.~\ref{fig:Ni_pes} (left) for various ellipsoidal shapes, {\it spherical}, {\it prolate}, {\it oblate} and in between (called {\it triaxial}).  The energy is obtained by the constraint Hartree-Fock (CHF) calculation with the same shell-model Hamiltonian as in Fig.~\ref{fig:Ni}.  The imposed constraints are given by the quadrupole moments in the intrinsic (body-fixed) frame.  This plot is usually called the Potential Energy Surface (PES).  The minimum energy occurs at the spherical shape (red sphere).  The constraints are changed to a more prolate deformed ellipsoid (blue object) along the upper-right axis (``prolate deformation'' in the figure).  The energy relative to the minimum energy climbs up by 6 MeV first.  This is because protons and neutrons must be excited across the magic gaps from the doubly closed shell in order to create states of deformed shapes (see Fig.~\ref{fig1}).  The energy then start to come down, as the quadrupole moments increase, thanks to the quadrupole interaction.  It is lowered by 3 MeV from the local peak to the local minimum.   Beyond the local-minimum area, the effect of the quadrupole interaction is saturated, and it cannot compete the energy needed for exciting more protons and neutrons across the gaps required by the constraints.  This energy variation appears as the basin in the 3-dimensional PES.  This is the usual explanation of the local deformed minimum.   The appearance of two (or more) different shapes with rather small energy difference is one of the phenomena frequently seen, and is called the shape coexistence  \cite{heyde2011}. 
The quadrupole interaction is undoubtedly among the essential factors of the shape coexistence.  But, this may not be a full story.


Figure~\ref{fig:Ni_pes} (right) exhibits the same energy along the axis lines of the left panel.  The positive (intrinsic) quadrupole moments imply prolate shapes (blue object in the left panel), whereas the negative ones oblate shapes (green object).  The red solid line shows the CHF results of the full Hamiltonian, whereas for the dashed line, the tensor monopole interactions between the neutron $g_{9/2}$ orbit and the proton $f_{5/2,7/2}$ orbits are practically removed.  This removal means no effects depicted in Fig.~\ref{fig:coexi} (d,e).
The dashed line displays a less pronounced prolate local minimum at weaker deformation with much higher excitation energy.  
The significant difference between the solid and dashed lines suggests that the monopole effects are crucial to lower this local minimum and stabilize it.
We now discuss the mechanism for this difference.
With the tensor monopole interaction, once sufficient neutrons are in $g_{9/2}$, the proton $f_{5/2}$-$f_{7/2}$ splitting is reduced, and this reduced splitting facilitates the mixing between these two orbits driven by the quadrupole interaction.  The resulting deformation is stronger compared to no tensor-force case. 
The tensor monopole interaction involving the neutron $g_{9/2}$ orbit produces extra binding energy, if more protons are in $f_{5/2}$ and less are in $f_{7/2}$.   This extra binding energy lowers the deformed states, otherwise they are high in energy because of the energy cost for promoting neutrons from the $pf$ shell to $g_{9/2}$.
Thus, a strong interplay emerges between the monopole interaction and the quadrupole interaction, and {\bf Type II shell evolution} materializes this interplay in the present case.  It enhances the deformation and lowers the energy of deformed states.  Without this interplay, as indicated by blue dashed line in Fig.~\ref{fig:Ni_pes} (right), the rotational  band corresponding the local minimum is pushed up by 2 MeV, and may be dissolved into the sea of many other states.  It is obvious that this interplay mechanism works self-consistently.   

\subsection{Tplot analysis}

\begin{figure}[bt]
\includegraphics[width=7.5 cm]{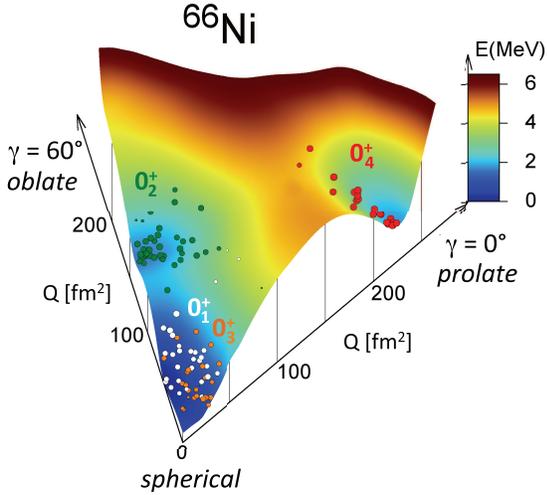}
\caption{PES and Tplot for $^{66}$Ni.  Taken from Fig. 1 of \cite{leoni_2017}. }
\label{fig:66Ni}
\end{figure}

The {\bf T-plot} was introduced in the same reference \cite{tsunoda2014}, so that what shapes are more relevant to individual eigenstates of the shell-model calculation. Let us take an example.  Figure~\ref{fig:66Ni} \cite{leoni_2017} depicts the PES of $^{66}$Ni with the same Hamiltonian as in Fig.~\ref{fig:Ni}.   The small circles on the PES are the T-plot.
The T-plot is obtained from MCSM eigenstate.  We therefore briefly explain the MCSM eigenstate.   An MCSM eigenstate, $\Psi$, is written, with the ortho-normalization, as
\begin{equation}
\label{eq:mcsm}
\Psi \, = \, \sum_k \, f_k \, \hat{{\mathcal P}}_{J^{\pi}} \, \phi_k \,\,, 
\end{equation}
where $f_k$ denotes amplitude, $\hat{{\mathcal P}}_{J^{\pi}}$ means the projection operator on to the spin/parity $J^{\pi}$ (this part is more complicated in practice), and $\phi_k$ stands for Slater determinant called ($k$-th) MCSM basis vector: $\phi_k$ = $\Pi_i \, c^{(k)\dagger}_i \,|0 \rangle$.  Here, $|0 \rangle$ is the inert core (closed shell), $c^{(k)\dagger}_i$ refers to a superposition of usual single-particle states, for instance, those of the HO potential,  
\begin{equation}
\label{eq:mcsm_bv} 
c^{(k)\dagger}_i \,=\, \sum_n \, D^{(k)}_{i,n} \, a^{\dagger}_n \,\, ,
\end{equation}
with $a^{\dagger}_n$ being the creation operator of a usual single-particle state, for instance, that of the HO potential, and $D^{(k)}_{i,n}$ denoting a matrix element.  By choosing a favorable matrix $D^{(k)}$, we can select $\phi_k$ so that such $\phi_k$ better contributes to the lowering of the corresponding energy eigenvalue.  Thus, the determination of $D^{(k)}$ is the core of the MCSM calculation. The index $k$ runs up to 50-100, but sometimes to 300 at maximum.  These are much smaller than the dimension of the many-body Hilbert space.


Each $\phi_k$ has intrinsic quadrupole moments ($\langle  \phi_k |\hat{Q}_0|  \phi_k \rangle $ and $\langle  \phi_k |\hat{Q}_2| \phi_k \rangle $), and its T-plot circle is placed according to those values on the PES with its area proportional to the overlap probability with the corresponding eigenstate, {\it i.e.} $\Psi$ in eq.~(\ref{eq:mcsm}).  Such T-plot circles are shown in Fig.~\ref{fig:66Ni}.  The white circles represent the MCSM basis vectors for the ground state, while the red circles indicate the MCSM basis vectors for the 0$^+_4$ state, which is strongly deformed.  Although there is no local minimum for oblate shape, the 0$^+_2$ state is shown to be moderately oblate deformed.
The T-plot can thus give partial labelling to fully correlated eigenstates for mean values as well as fluctuations with respect to their quadrupole shapes.   The advantages of mean-field approaches are now nicely incorporated into the shell model.  
    
\subsection{Short summary of Sec. \ref{typeII}}

Type II shell evolution occurs in various cases, especially in a number of shape coexistence cases, providing deformed states with stronger deformation, lower excitation energies and more stabilities.  It is an appearance of the monopole-quadrupole interplay, and plays crucial roles in various phenomena including the first order phase transition (Zr isotopes \cite{togashi_2016,kremer_2016,singh_2018}), the second order phase transition (Sn isotopes \cite{togashi_2018}), the multiple even-odd phase transitions (Hg isotopes \cite{marsh_2018}) as well as the raising of the intruder band due to the suppression of the type II shell evolution (lighter Ni isotopes \cite{leoni_2017,marginean_2020}).   As the involvement of the monopole interaction in this manner had not been recognized, {\bf Type II shell evolution appears to be among the emerging concepts} of nuclear structure.    
The Type II shell evolution has been clarified by the T-plot in many cases.  Including other contributions, {\bf the T-plot is undoubtedly one of the emerging concepts of nuclear structure}, apart from its impact on the computational methodology.

\section{Self-organization and collective bands in heavy nuclei \label{coll}}

We now proceed to more general cases of the monopole-quadrupole interplay.  
This interplay leads to unexpected consequences in the under-lying mechanism of collective bands of heavy nuclei \cite{otsuka_2019}, beyond the standard textbooks.  

The MCSM has become powerful enough \cite{mcsm4} to reproduce collective bands of heavy nuclei such as $^{154}$Sm and $^{166}$Er, with one and half HO major shells \cite{otsuka_2019}.  We sketch the new findings by using the results of such most advanced MCSM calculations.

\subsection{Shape coexistence in $^{154}$Sm}
\begin{figure*}[bt]
\includegraphics[width=10.46 cm]{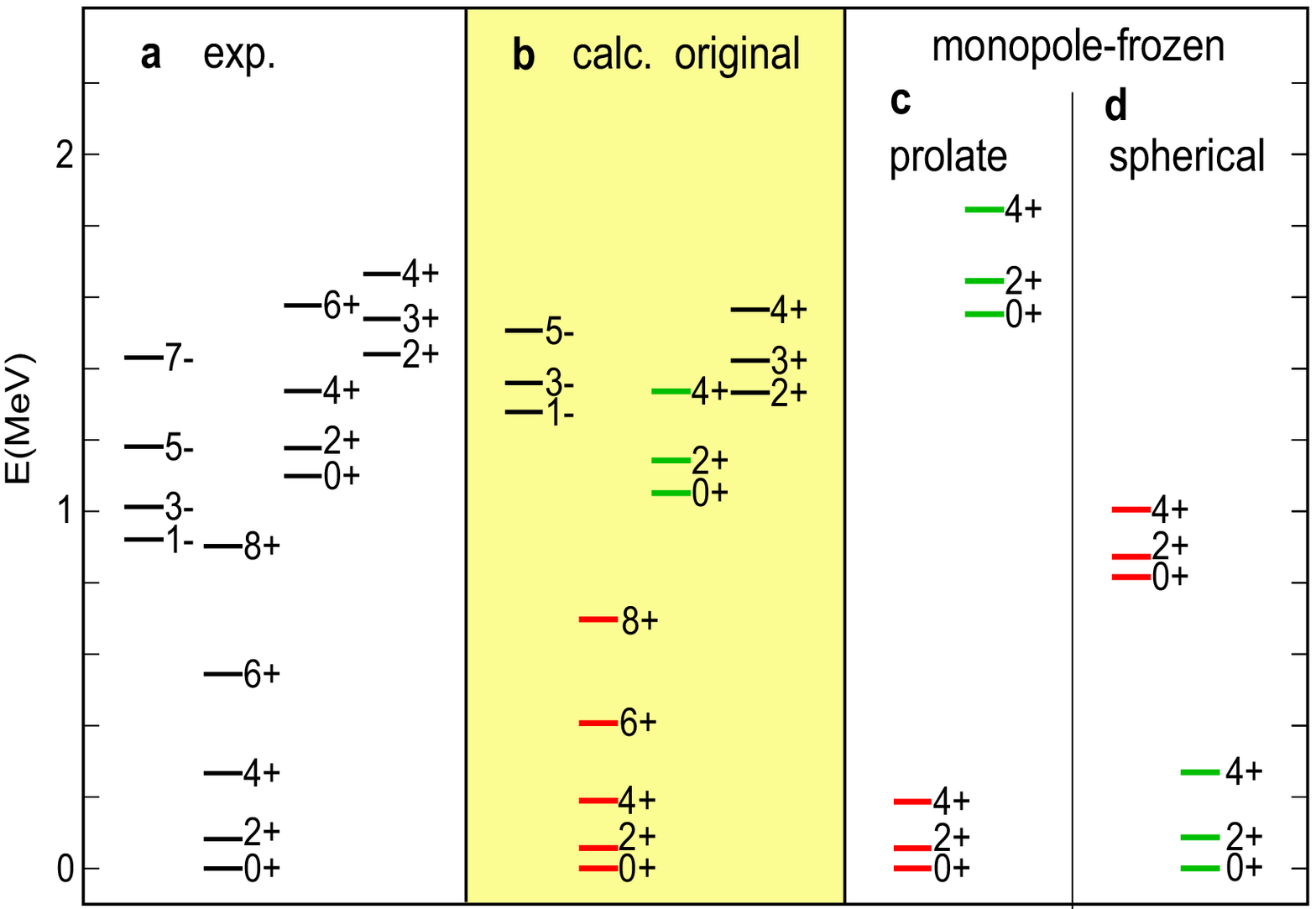}
\includegraphics[width=6.37 cm]{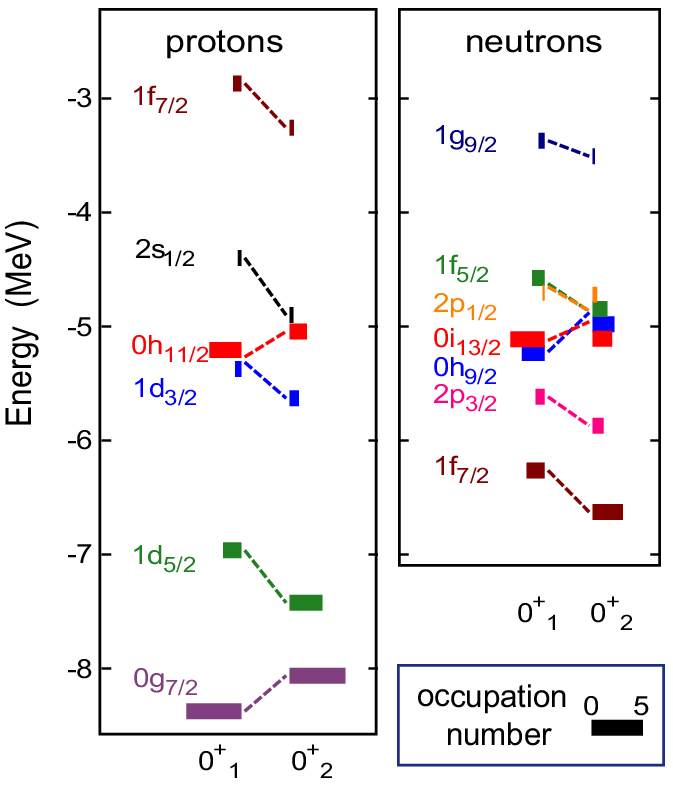}
\caption{(left) Experimental \cite{ensdf}, calculated original and monopole-frozen energy levels of $^{154}$Sm.  
(right) ESPE (vertical position) and occupation number
(horizontal width).
Taken from Figs. 2 and 3 of \cite{otsuka_2019}. }
\label{fig:Sm}
\end{figure*}

Figure~\ref{fig:Sm} shows low-lying energy levels of $^{154}$Sm.  The present MCSM calculation can describe the four low-lying bands including negative-parity one.  The agreement between the theoretical levels in panel {\bf a}  and the experimental levels in panel {\bf b} is rather good.  Although the importance of the quadrupole interaction is evident for the formation of deformed rotational bands, one can investigate to what extent the monopole interaction is involved.  The monopole interaction here was obtained from the shell-model interactions, comprising  the central, tensor and other components.     


The monopole interaction is an operator, but we ``freeze'' it now: its ESPE expectation values $\langle \hat{\epsilon}^{p,n}_{j} \rangle$ are calculated for the state to be specified, and the obtained values are adopted as the SPEs, $\epsilon^{p,n}_{0;j}$, with the monopole interaction removed.  We then perform the shell model calculation and draw the PES.  This toy game is called the ``monopole frozen'' analysis \cite{otsuka_2019}, as the monopole properties are included only through the specified state.  Panel {\bf c} of Fig.~\ref{fig:Sm} exhibits the energy levels obtained by the monopole-frozen analysis referring to the ground state.  The band built on the 0$^+_2$ state (often called the $\beta$ band) is lifted up by 0.5 MeV ($\sim$50\% of the original excitation energy), suggesting that the active monopole interaction produces a substantial lowering of this state.  Panel {\bf d} shows the monopole-frozen analysis referring to the spherical HF state: the ground state is no longer prolate, but triaxial, with the wave function close to the 0$^+_2$ state of the original Hamiltonian.  Thus, the crucial effect of the monopole interaction is verified.

Right panel of Fig.~\ref{fig:Sm} shows the actual values of $\langle \hat{\epsilon}^{p,n}_{j} \rangle$ for the 0$^+_1$ and  0$^+_2$ states.  This figure demonstrates the significant  differences between two sets of the ESPE values.  The occupation numbers are also different: there are more half-filled orbits for the 0$^+_2$ state, which is indicative of its triaxial nature.   The smaller occupation numbers of unique-parity orbits are also consistent with the tendency away from the prolate shape.

\begin{figure*}[bt]
\includegraphics[width=16 cm]{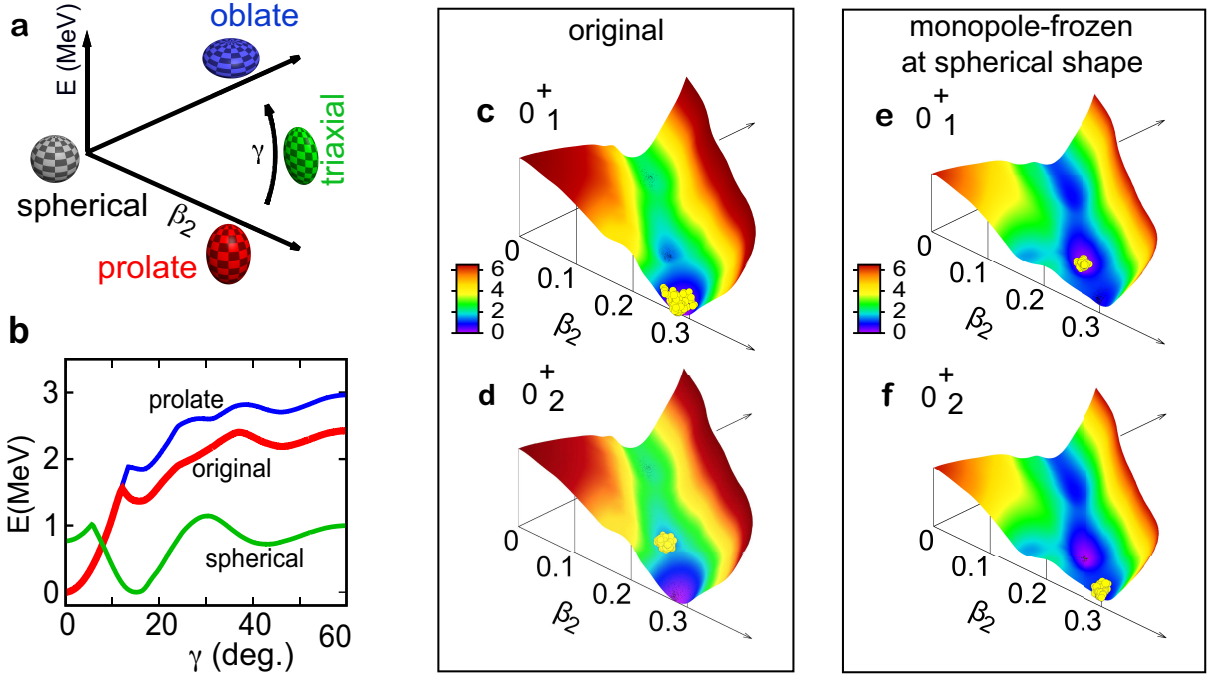}
\caption{Properties of the 0$^+_{1,2}$ states of $^{154}$Sm. (a) Deformation
parameters and shapes. (b) Lowest values of PES for a given $\gamma$ value
for the original, prolate and spherical monopole-frozen
cases. (c–f) Three-dimensional T-plot in the original or
spherical monopole-frozen case.
Based on Fig. 3 of \cite{otsuka_2019}. }
\label{fig:Sm_PES}
\end{figure*}

Figure~\ref{fig:Sm_PES} {\bf c} and {\bf d} show the T-plot for the original interaction, while Fig.~\ref{fig:Sm_PES} {\bf e} and {\bf f} depict the T-plot for the the monopole-frozen interaction obtained with the spherical HF state.  The T-plot patterns are consistent with the above features suggested by the shell-model diagonalization.  The cut of the PES shown in panel {\bf b} suggests that the local minimum is raised by the monopole-frozen process with  the ground state.  

Panels {\bf c} and {\bf d} depict a valley of  the PES with a local minimum around $\gamma = 15 ^{\circ}$.   Similar valleys are seen in the PES obtained by the mean-field calculations \cite{robeldo_2008,li_2009}, implying that this valley likely has a common origin.   On the other hand, one can state that the present monopole effect results in the local minimum, not simply the valley.   It is of interest to refine the monopole interaction in mean-field models.   

Regarding the so-called $\beta$ vibration picture of the  the 0$^+_2$ state, the present view  is opposed to such a conventional view.   
The triaxial deformation is shared by the members not only of the 0$^+_2$ band but also of the 2$^+_3$ band (usually called $\gamma$ band), as can be verified by their T-plots.  Namely, the 0$^+_2$ state is the ``ground'' state of the triaxial states to which both the 0$^+_2$ and 2$^+_3$ bands belong.  In short, this is a shape coexistence between the prolate and triaxial shapes assisted by the interplay between the monopole interaction and 
the quadrupole deformation.

\subsection{Collective bands and $\gamma$ vibration in $^{166}$Er \label{166Er}}

\begin{figure*}[bt]
\includegraphics[width=15 cm]{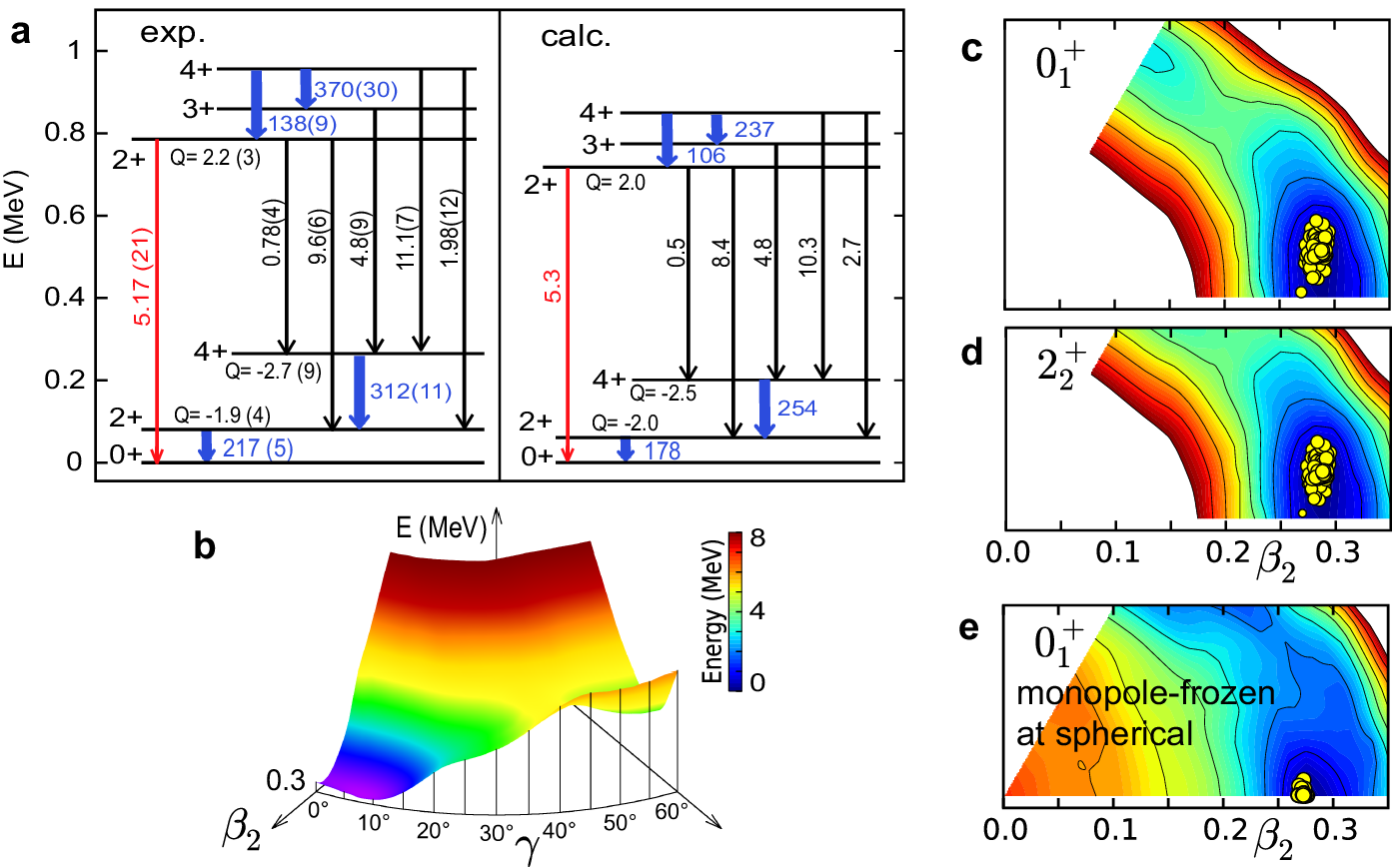}
\caption{
Experimental and calculated properties of the lowest
states of $^{166}$Er. {\bf a} Energy levels and electromagnetic transitions
(W.u.) \cite{ensdf} as well as spectroscopic electric quadrupole moments
(eb) \cite{stone2016}. {\bf b} Three-dimensional PES and its cut surface for
$\beta_2$= 0.3. {\bf c-e} T-plots for the 0$^+_1$ and 2$^+_2$ states, and for the
monopole-frozen 0$^+_1$ state at spherical shape.
Based on Fig. 4 of \cite{otsuka_2019}. }
\label{fig:Er}
\end{figure*}

The features of the collective motion in $^{166}$Er have been studied by the MCSM similarly well (see Fig.~\ref{fig:Er} {\bf a}).  Among so-called rotational nuclei, $^{166}$Er is characterized by particularly low-lying 2$^+_2$ state and the so-called $\gamma$ band built on it.  Aage Bohr stressed that this 2$^+_2$ state was a $\gamma$ vibration from the prolate ground state \cite{bohr1952,bohr_mottelson1953,bohr_mottelson_book2}.   The relatively strong 2$^+_2$$\rightarrow$ 0$^+_1$ E2 transition (B(E2)$\sim$ 5 e fm$^2$, see panel {\bf a}) was ascribed to  the annihilation of one $\gamma$ phonon in the 2$^+_2$ state.  This was one of the major points of the Nobel lecture by Aage Bohr, and has been a common sense as stated in many textbooks of nuclear physics.  We now challenge this traditional belief, by utilizing the recent MCSM calculation.  It is reminded that no firm experimental evidence to uniquely pin down the $\gamma$-vibration nature of $^{166}$Er has been reported, and also that in a systematic calculation of many heavy nuclei \cite{delaroche_2010}, the excitation energies of the 2$^+_2$ states in the $\gamma$ band appeared to be about twice higher than the observed values, despite much better description of those of the 2$^+_1$ state in the ground band. 
 
Figure~\ref{fig:Er} {\bf b} shows the calculated PES, which shows the minimum not at $\gamma =0 ^{\circ}$ but around $\gamma =9 ^{\circ}$ (see also \cite{tsunoda_2021}).
The T-plot is shown for the $0^+_1$ and $2^+_2$ states in panels {\bf c} and {\bf d}, respectively.  The patterns of the T-plot circles are nearly identical between these two panels.   This is consistent with a (rigid) triaxial interpretation, and indeed E2 transition strengths follow the predictions of Davidov triaxial model \cite{davydov1,davydov2} with $\gamma =9 ^{\circ}$.  Certainly, a pure rigid triaxiality is not the correct picture, and there are quantum fluctuations, as evident from panels {\bf c} and {\bf d} \cite{tsunoda_2021}.   After all, the displacement from the $\gamma =0 ^{\circ}$ is ovbious.  Note that the so-called $K^{\pi}$=4$^+$  double $\gamma$-phonon states are also described well by the present work \cite{tsunoda_2021}, whereas its reproduction within the phonon model is known to be very difficult.
The triaxiality of $^{166}$Er is also suggested by the triaxial projected shell model, although the rigid-triaxiality is not an outcome but an assumption \cite{sun2000,sun2002}.

The monopole-frozen analysis referring to the spherical CHF state shows that the ground state moves to $\gamma =0 ^{\circ}$, confirming the important role of the monopole interaction activated.   The triaxial ground states are now shown to appear in a large number of nuclei in the nuclear chart, besides the known triaxial domain \cite{hayashi_1984}. 

\subsection{A historical touch and a short summary of Sec. \ref{coll}}

The collective bands in heavy nuclei have traditionally been understood in terms of the ground band with axially symmetric prolate shape and the side bands with the $\beta$ or $\gamma$ vibrational excitations from the ground state.  This picture is consistent with the Nilsson model \cite{nilsson1955}, and was confirmed by the Pairing + Quadrupole-Quadrupole (P+QQ) model \cite{kumar1968,bes1969}, where the monopole interaction is not included, however.  
It has been shown in this section that the monopole interaction is crucial also for the collective bands in heavy nuclei.  We just note that in lighter nuclei, the situation can be different mainly because of small model spaces comprising single or a few active orbits, where the rotational motion has been nicely described by symmetry-based approaches, {\it e.g.} SU(3) model of Elliott for the $sd$ shell \cite{elliott1958a,elliott1958b}, and by realistic calculations, {\it e.g.} on $^{48}$Cr \cite{caurier1994}.  

Regarding heavy nuclei, for individual rotational bands, the monopole interaction contributes differently, and the intrinsic structure is determined not only by the quadrupole interaction but also by the monopole interaction, as verified by the monopole-frozen analyses.  Thus, the monopole-quadrupole interplay arises.  The monopole interaction does not directly drive the deformation, but optimizes the ESPEs so that more binding energy is gained.  This gain is state-dependent, and even can alter the ordering of bands as mentioned above.
The present monopole-quadrupole interplay can be described also from the viewpoint of the self-organization \cite{otsuka_2019}: 
the nucleus is changed from a disorder (original SPEs) to an order (ESPEs tailored to the shape of interest) by activating the monopole interaction.
As this occurs ``purposely'' towards certain shapes with positive feedback particularly between the monopole and quadrupole effects, the whole picture fits well the (quantal) self-organization \cite{otsuka_2019}.
{\bf The self-organization for collective bands is among the emerging concepts of nuclear structure,} showing novel consequences.  For example, {\bf the dominant fraction of the ground states of heavy nuclei are expected to show triaxial shapes, as another emerging concept of nuclear structure,} in contrast to the traditional view of the prolate shape dominance in those states.

Appendix \ref{A2} presents a possible extension or generalization of the current idea to  ``many-ingredient'' systems outside nuclear physics.

\section{Dripline mechanism \label{drip}}

\subsection{Traditional view}

\begin{figure*}[bt]
\includegraphics[width=17 cm]{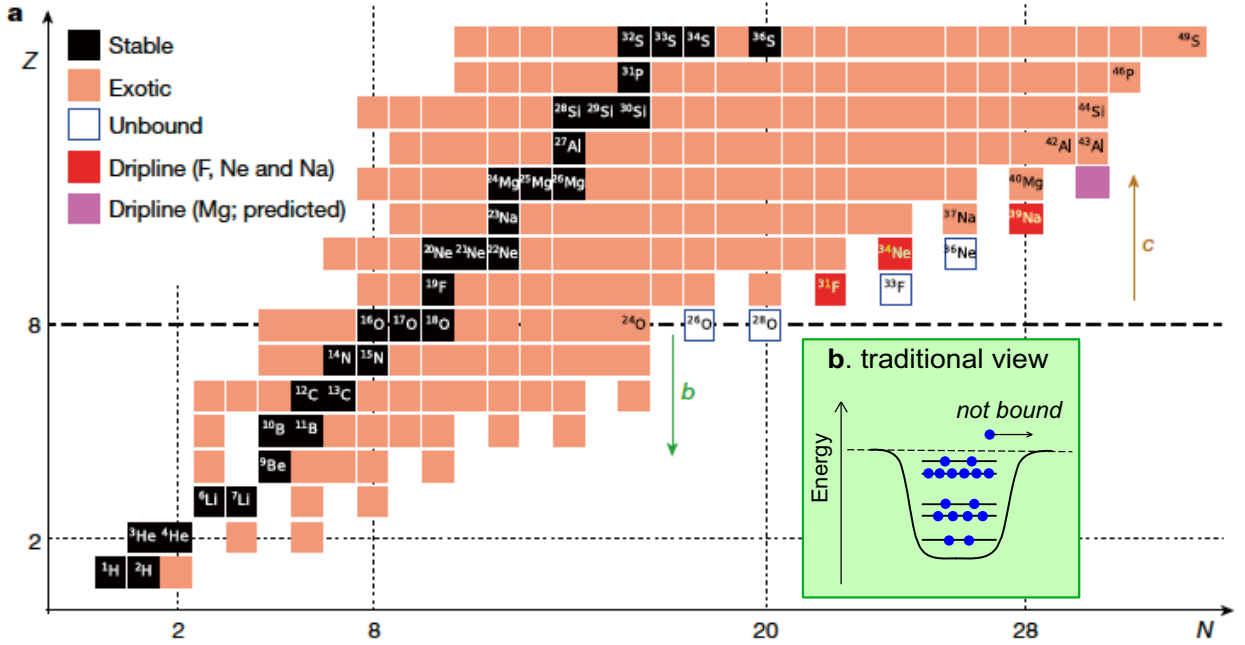}
\caption{Part of the nuclear chart and schematic illustrations of the traditional view of the dripline.  Based on Fig. 2 of \cite{dripline2020}. }
\label{fig:nchart}
\end{figure*}

Figure~\ref{fig:nchart} {\bf a} shows the lower-left part of the nuclear chart (Segr\`e chart) for  $Z\le16$.   
The black squares represent stable nuclei, while the orange ones exotic nuclei (see Sec.~\ref{intro}).   An isotopic chain is a horizontal belt, and its neutron-rich end is called neutron dripline.  The location of the dripline in the nuclear chart implies the extent of the isotopes, and is of fundamental importance to nuclear science.  The experimental determination of the dripline is a very difficult task.  Very recently, as shown by red squares in Fig.~\ref{fig:nchart}  {\bf a}, the driplines of F and Ne isotopes and its candidate of Na isotope were reported \cite{ahn2019}.  

\begin{figure*}[bt]
\includegraphics[width=13 cm]{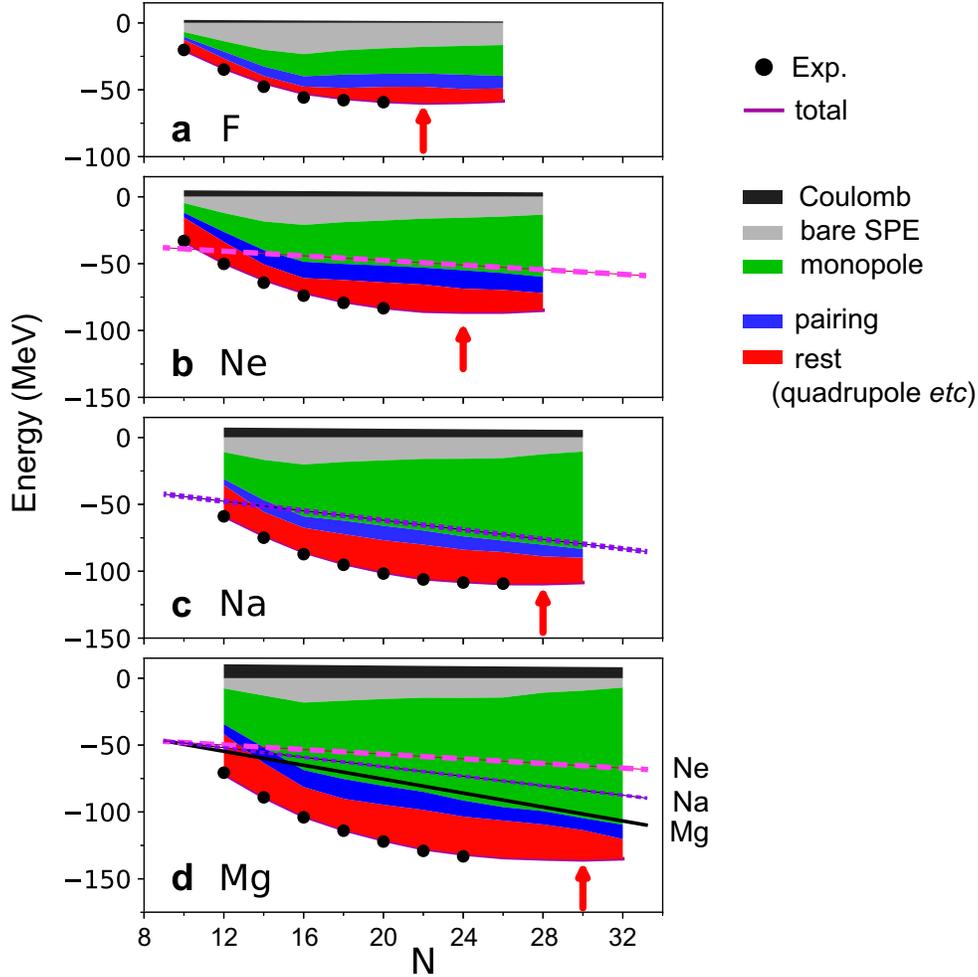}
\caption{
Ground-state energies of even-N isotopes of F, Ne, Na and Mg,
relative to the $^{16}$O value. Colored segments exhibit decompositions into
various effects from the monopole (green), pairing (blue) and rest (such as
quadrupole) (red) components of the effective nucleon–nucleon interaction as
well as those from Coulomb interaction (black) and single-particle energies
(bare SPE; grey). The monopole effect grows steadily as a function of $N$ in all
cases, as highlighted by straight lines: dashed (Ne), dotted (Na) and solid (Mg).  The
experimental values are indicated by black circles \cite{ensdf}. The theoretical driplines indicated by red arrows. 
Modified from Fig. 4 of \cite{dripline2020}. }
\label{fig:drip_mono}
\end{figure*}   

The traditional view of the dripline is shown in Fig.~\ref{fig:nchart} {\bf b}: all bound single-particle orbits are occupied, and the next neutron goes away.  It is an open question whether this view is valid for all nuclei or not.  We look into this question now \cite{dripline2020}.
 
The structure of neutron-rich exotic isotopes of F, Ne, Na and Mg can be well described by the shell model calculation with the full $sd$+$pf$ shells and the EEdf1 interaction \cite{n_tsunoda2017}.  This interaction was derived from the chiral EFT interaction of Machleidt and Entem \cite{machleidt2011}, first processed by the V$_{{\rm low}-k}$ method \cite{bogner_2002,nogga_2004}, and then processed by the EKK (Extended Krenciglowa-Kuo) method \cite{takayanagi2011a,takayanagi2011b,n_tsunoda2014}.  The V$_{{\rm low}-k}$ method is used to transform the nuclear forces in the free space into a tractable form for further treatments.  The V$_{{\rm low}-k}$ method has been adopted for the derivation of other modern shell-model interactions, for instance, the one by Corraggio {\it et al.}, for Sn and Cr-Fe regions \cite{coraggio2009,arnswald2017}. 

The present work is unique in the usage of  the EKK method, which is an extension of the many-body perturbation theory (MBPT) \cite{jensen_1995}.  The MBPT produced the G-matrix interactions in its early formulations \cite{kb1966}, from which many useful shell-model interactions have been constructed (see subsect. \ref{subsec:vmu}).  However, the G-matrix formulation has a limitation that if two major shells are merged, the results may diverge \cite{n_tsunoda2014}.  As the gap between two shells often vanishes or becomes smaller in exotic nuclei, this difficulty can be fatal there.  The EKK method nicely avoids this difficulty besides other merits.   
 
The EEdf1 interaction has thus been derived in an {\it ab initio} way by the V$_{{\rm low}-k}$ and EKK methods, and describes the properties of the ground and low-lying states of F, Ne, Na and Mg isotopes quite well \cite{n_tsunoda2017,dripline2020}.

\subsection{Monopole-quadrupole interplay for the driplines}


\begin{figure*}[bt]
\includegraphics[width=17.5 cm]{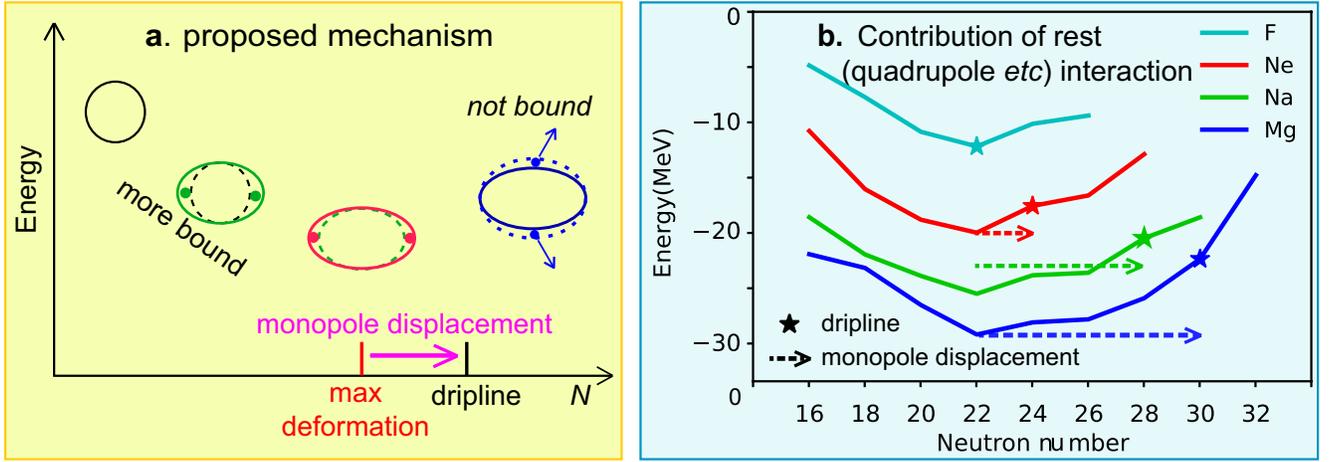}
\caption{{\bf a} Presently proposed mechanism based on shape evolution and the resulting change of the binding energy.
{\bf b} The rest-term contribution to the ground-state energies for F, Ne, Na and Mg isotopes. Dashed arrows indicate the monopole displacement.
Modified from Figs. 2 and 6 of \cite{dripline2020}. }
\label{fig:drip_shift}
\end{figure*}   

Figure~\ref{fig:drip_mono} shows the ground-state energies of F, Ne, Na and Mg isotopes as functions of the neutron number $N$.  These energies are decomposed into several pieces according to their origins: SPE (on top of the $^{16}$O inert core), monopole, pairing and rest terms. The Coulomb contribution is ignored in the following discussion, because of virtually no relevance. Here, the multipole interaction is divided into the pairing and rest terms.  The pairing is the BCS-type pairing interaction acting on two neutrons coupled to $J^{\pi}=0^+$ and on two protons coupled to $J^{\pi}=0^+$.  The rest term means the multipole interaction subtracted by the pairing term.  Although the rest term contains many different pieces, its major effects in the present discussion is simulated by the quadrupole interaction.  This is the reason why the rest term is associated with "(quadrupole etc)" in the figure. 

The lower edges of the red areas exhibit the ground-state energies as functions of the neutron number $N$, while only even $N$ values are taken.  These values show a good agreement with measured values shown by black dots.  As long as the ground-state energy becomes lower as $N$ increases, the isotope gains more binding energy by having more   neutrons, and the isotope chain is stretched.  However, if the ground-state energy is not lowered, there is no gain of the binding energy by having these extra neutrons; these extra neutrons are emitted, and the neutron dripline implies the nucleus with the lowest ground-state energy.  The driplines obtained by the present calculation are shown by red arrows for each isotopic chain, reproducing experimental driplines for F, Ne and Na isotopes \cite{ahn2019}. 

We focus on the lower edge of the green areas in Fig.~\ref{fig:drip_mono}.  This represents the monopole contributions comprising the SPE and the monopole interaction.     
For Ne, Na and Mg isotopes, this edge is lowered almost linearly as $N$ increases from  $N$=16 to each dripline.  We then fit the edge with pink dashed, purple dotted and black solid lines for F, Ne and Mg isotopes, respectively.   The lines of Ne and Na isotopes are copied to the panel for Mg, with their positions adjusted at a certain $N$.  It is evident that the lines become steeper almost linearly as $Z$ increases. This edge is almost flat for F isotopes for $N\ge16$, and this feature will be discussed later.

Figure~\ref{fig:drip_mono} indicates that the effect of the pairing term shows small variations.
In contrast, the rest terms changes more, and are largely due to the quadrupole interaction.   Figure~\ref{fig:drip_shift}(a) schematically indicates the variation of the effect of the quadrupole interaction: The effect is small at the far left position with a spherical shape.  As some neutrons are added, the shape is deformed, and the ground-state energy is lowered due to the quadrupole interaction.  This trend continues, but becomes its maximum at a certain value of $N$ (red object in the figure).   
However, the dripline is not determined just by this maximum point. 

Figure~\ref{fig:drip_shift}(b) depicts the actual effect of the rest term.
It follows the trend illustrated in Fig.~\ref{fig:drip_shift}(a), with the maximum effect at $N$=22 in all four chains.  However, the driplines are different among these four.  This is due to the monopole interaction.  Let me explain it, by taking the Mg isotopes as an example.    
The black straight line of the monopole effect in Fig.~\ref{fig:drip_mono}(d) depicts about 3 MeV lowering per additional neutron, implying about 6 MeV for additional two neutrons.  After $N$=22, the rest effect loses its magnitude.  If the loss is less than the monopole gain ($\sim$6 MeV), this loss is compensated by the monopole effect.  However, the loss becomes larger for $N$ larger, and at a certain point, the loss exceeds the monopole compensation.  The dripline thus arises with the ``monopole displacement'' from $N$=22 to $N$=30 as shown in Fig.~\ref{fig:drip_shift}(b) (and also in Fig.~\ref{fig:drip_shift}(a) schematically).  

The monopole effect depends directly on the number of protons, as visualized by three straight lines Fig.~\ref{fig:drip_mono}.  Consequently, the monopole displacement is $\Delta N$=2 (6)  for Ne (Na) isotopes.  For F isotopes, the monopole effect is negligibly small for $N \ge$16, and the dripline is located at the maximum rest (quadrupole {\it etc}) effect.  

\subsection{Stability of spherical isotopes and the monopole-quadrupole interplay}

An immediate lemma of the present dripline mechanism is that the driplines of spherical nuclei, such as Ca, Sn and Pb isotopes, may be further away from the stability line than other elements.  One can assume a basically constant pairing contribution and a minor rest-term contribution.  These two are thus irrelevant to the driplines of these isotopes.  The remaining monopole effect gradually changes, pushing the driplines away.

\subsection{A short summary of Sec. \ref{drip}}

The {\bf present new dripline mechanism \cite{dripline2020} involves the monopole-quadrupole interplay}, and is {\bf one of the emerging concepts}.  It definitely differs from the traditional mechanism of the single-particle origin, where neutron halo arises at extremes \cite{tanihata1985,hansen1987}.  In the new mechanism, the coupling to continuum may be visible if the monopole effect vanishes like heavy F isotopes \cite{dripline2020}. 
As $Z$ changes, two dripline mechanisms may appear alternatively, but the present one may be more relevant to heavier nuclei where the deformation develops more.

\section{Prospect \label{prospect}}

As this article is a kind of summary, I am afraid that a summary section may be redundant.   I state some prospects.  First of all, {\it ab initio} no-core Monte Carlo Shell Model calculations became feasible recently up to $^{12}$C and beyond \cite{abe2021}, and as an example, we can look into $\alpha$ clustering in light nuclei, {\it e.g.} the Hoyle state, with correlations produced by nuclear forces \cite{otsuka2022}.  This direction will produce a major outcome from the shell model.  This includes clarifications of $\alpha$ decay, $\alpha$ knockout, {\it etc.}  Another major frontier is the quest for fission dynamics and superheavy elements, with (almost) full inclusion of the correlations due to nuclear forces.   

Although more computer powers and further advancements in computational methodology are needed also, the perspectives of the shell model look unlimited, to me.  {\it May the (nuclear) force be with you.} 

\vspace{6pt} 

\section*{Acknowledgements}
The author is grateful to Drs. A. Gargano and S.M. Lenzi for their interests and for the invitation to this valuable program.  
He acknowledges Drs. T. Abe, Y. Akaishi, B.A. Brown, P. Van Duppen, B. Fornal, R. Fujimoto, A. Gade, H. Grawe, R.V.F. Janssens, M. H.-Jensen, K. Heyde, J. Holt, M. Honma, M. Huyse, Y. Ichikawa, S. Leoni, T. Mizusaki, N. Pietralla, P. Ring, E. Sahin, J. P. Schiffer, A. Schwenk, N. Shimizu, O. Sorlin, Y. Sun, T. Suzuki, K. Takayanagi, T. Togashi, K. Tsukiyama, N. Tsunoda, Y. Tsunoda, Y. Utsuno, S. Yoshida, and H. Ueno for their valuable direct collaborations towards the works presented in this article, and thanks many others for their useful comments and helps.  
The comment by Dr. Y. Tsunoda on Appendix \ref{A1} is appreciated.
This work was supported in part by MEXT 
as "Program for Promoting Researches on the Super computer Fugaku" (Simulation for basic science: from fundamental laws of particles to creation of nuclei) and by JICFuS. This work was supported by JSPS KAKENHI Grant Numbers JP19H0514, JP21H00117.\\

\vspace{1cm}

\appendix
\section{Note on the relation between the present ESPE and the Baranger's ESPE
\label{A1}}
This is a short note on the the relation between the present ESPE and Baranger's ESPE \cite{baranger1970} discussed in \cite{otsuka2020}.  A possible problem was pointed out by Y. Tsunoda.  Although the relevant arguments and results in \cite{otsuka2020} are basically correct, the following term is found to be added to eq.(43) of \cite{otsuka2020}:
$- 1/(2j+1) V^m (j, j) \langle 0  | \hat{n}_{j} \, | 0  \rangle,\,$  
where $j$ includes the index, proton or neutron.  So, this is the contribution from the interaction between a neutron orbit $j$ and the same neutron orbit $j$ (or between protons similarly), of which the monopole interaction is known to be weak.  In addition, the factor 1/(2j+1) reduces this quantity.  Because of all these factors combined, the correction is quite minor.   This correction does not change the basic equivalence relation between the two schemes.

 \section{Self-organization and its extension to other ``many-body'' systems \label{A2}}
We here discuss briefly how the present self-organization mechanism may be applied to other systems comprising many constituents, including human societies.  One of the essential points is two interactions with different characters: one drives the system into specific modes, as denoted by the mode-driving force.  The mode here generally refers to a collective phenomenon involving many constituents, like the  shape of an atomic nucleus.  A certain resistance usually exists against the mode development.  The other interaction is to control the resistance, called the resistance-control force.   The monopole interaction in this work is an example.  The resistance-control force does not create any mode, being neutral.  However, it can change the disorder in the original environment (=original SPE in this work) to the order where the resistance is weakened for certain modes (ESPE tailored to the shape). This order thus gives extra stability to the system, to varying degrees depending on the modes.  Thus, the resistance-control force can be a crucial factor in determining which mode gains the maximum stability ({\it i.e.} binding energy).  Obviously, in many systems, only the maximum-stability mode matters, which may not be the one most favored by the driving force.      
If this general idea can be applied to various problems, including social/economical issues, it is of great interest.  While the mode varies over different systems, the mode-driving force may be visible.  The resistance-control force, however, may not be so, because it exhibits less characteristics (like the monopole interaction in atomic nuclei).  Studies in this direction can be of interest.  What are the resistance and its control force in human societies ?


\begin{thebibliography}{999}
\bibitem{mayer1949}
Goeppert Mayer, M. 
On closed shells in nuclei. II.
{\em Phys. Rev.} {\bf 1949}, {\em 75}, 1969.

\bibitem{haxel1949}
Haxel, O.; Jensen, J. H. D.; Suess, H. E. 
On the ``magic numbers'' in nuclear structure.
{\em Phys. Rev.} {\bf 1949}, {\em 75}, 1766.

\bibitem{talmi1962} 
Talmi, I.  
Effective Interactions and Coupling Schemes in Nuclei.
{\em Rev. Mod. Phys.} {\bf 1962}, {\em 34}, 704-722.

\bibitem{caurier2005}
Caurier, E.; Martínez-Pinedo, G.; Nowacki, F.; Poves, A.; Zuker, A. P.
The shell model as a unified view of nuclear structure. 
{\em Rev. Mod. Phys.} {\bf 2005}, {\em 77}, 427–488.

\bibitem{gade2008}
Gade, A.; Glasmacher, T. 
In-beam nuclear spectroscopy of bound states with fast exotic ion beams.
{\em Prog. Part. Nucl. Phys.} {\bf 2008}, {\em 60}, 161-224.

\bibitem{sorlin2008}
Sorlin, O.; Porquet, M.-G. 
Nuclear magic numbers: New features far from stability.
{\em Prog. Part. Nucl. Phys.} {\bf 2008}, {\em 61}, 602-673.

\bibitem{nakamura2017}
Nakamura, T.; Sakurai, H.; Watanabe, H.
Exotic nuclei explored at in-flight separators.
{\em Prog. Part. Nucl. Phys.} {\bf 2017}, {\em 97}, 53-122.

\bibitem{rainwater1950}
Rainwater, J. 
Nuclear energy level argument for a spheroidal nuclear model.
{\em Phys. Rev.} {\bf 1950}, {\em 79}, 432 (1950).

\bibitem{bohr1952}
Bohr, A.
The Coupling of Nuclear Surface Oscillations to the Motion of Individual Nucleons.
{\em Mat. Fys. Medd. Dan. Vid. Selsk.} {\bf 1952}, {\em 26}, no. 14;
Bohr, A.
Rotational Motion in Nuclei.
In {\em Nobel Lectures, Physics 1971—1980}, 
Lundqvist S., Ed.; World Scientific: Singapore, 1992; pp. 213--232;
https://www.nobelprize.org/prizes/physics/1975/bohr/facts/.

\bibitem{bohr_mottelson1953}
Bohr, A.,; Mottelson, B.R. 
Collective and Individual-Particle Aspects of Nuclear Structure.
{\em Mat. Fys. Medd. Dan. Vid. Selsk.} {\bf 1953}, {\em 27}, no. 16.

\bibitem{BMbook1}
Bohr, A.; Mottelson, B.R. \textit{Nuclear Structure I}; Benjamin: New York, USA, 1969.

\bibitem{bohr_mottelson_book2}
Bohr, A.; Mottelson, B.R. \textit{Nuclear Structure II}; Benjamin: New York, USA, 1975.

\bibitem{schaefer2014}
Sch\"afer, T. 
Fermi liquid theory: A brief survey in memory of Gerald E. Brown, 
{\em Nucl. Phys. A} {\bf 2014}, {\em 928}, 180-189.

\bibitem{otsuka2020} 
Otsuka, T.;  Gade, A.; Sorlin, O.; Suzuki, T.; Utsuno, Y.
Evolution of shell structure in exotic nuclei. 
{\em Rev. Mod. Phys.} {\bf 2020}, {\em 92}, 015002.

\bibitem{ragnarsson1995}
Ragnarsson, I.; Nilsson, S. 
in {\it Shapes and Shells in Nuclear Structure}; 
Cambridge University Press, Cambridge, UK, 1995.

\bibitem{poves1981}    
Poves, A.; Zuker, A.
Theoretical Spectroscopy and the fp shell.  
{\em Phys. Rep.} {\bf 1981}, {\em 70}, 235-314.

\bibitem{bansal1964}
Bansal, R.K.; French, J.B. 
Even-parity-hole states in f7/2-shell nuclei.
{\em Phys. Lett.} {\bf 1964}, {\em 11}, 145-148.

\bibitem{baranger1970}
Baranger, M.
A definition of the single-nucleon potential, 
{\em Nucl. Phys. A} {\bf 1970}, {\em 149}, 225.

\bibitem{storm1983}
Storm, M.; Watt, A.; Whitehead, R. 
Crossing of single-particle energy levels resulting from neutron excess in the sd shell
{\em J. Phys. G} {\bf 1983}, {\em 9}, L165-168.

\bibitem{FP1977}
Federman, P.; Pittel, S.
Towards a unified microscopic description of nuclear deformation.
{\em Phys. Lett. B} {\bf 1977}, {\em 69}, 385-388.

\bibitem{grawe2004}
Grawe, H. 
Shell model from a practitioner’s point of view. 
In {\em The Euroschool Lectures on Physics with Exotic Beams, Vol. I};
Al-Khalili, J.; Roeckl, E., Eds.; Springer, Berlin/Heidelberg, Germany, 2004; pp. 33--75.

\bibitem{otsuka_2005}
Otsuka, T.; Suzuki, T.; Fujimoto, R.; Grawe, H.; Akaishi, Y. 
Evolution of the nuclear shells due to the tensor force.
{\em Phys. Rev. Lett.} {\bf 2005}, {\em 95}, 232502.

\bibitem{otsuka_2010}
Otsuka, T.; Suzuki, T.; Honma, M.; Utsuno, Y.; Tsunoda, N.; Tsukiyama, K.; Hjorth-Jensen, M.
Novel Features of Nuclear Forces and Shell Evolution in Exotic Nuclei.
{\em Phys. Rev. Lett.} {\bf 2010}, {\em 104}, 012501.

\bibitem{usd}
Brown, B.A.; Wildenthal, B.H. 
Status of the nuclear shell model.
{\em Annu. Rev. Nucl. Part. Sci.} {\bf 1988}, {\bf 38}, 29-66.

\bibitem{gxpf1a}
Honma, M.; Otsuka, T.; Brown, B.A.; Mizusaki, T.
Effective interaction for pf-shell nuclei.
{\em Phys. Rev. C} {\bf 2002}, {\em 65}, 061301(R);
New effective interaction for pf-shell nuclei and its implications for the stability of the N=Z=28 closed core.
{\em Phys. Rev. C}  {\bf 2004}, {\em 69}, 034335.

\bibitem{kb1966}
Kuo, T.T.S. and Brown, G.E.,
Structure of finite nuclei and the free nucleon-nucleon interaction An application to $^{18}$O and $^{18}$F. 
{\em Nucl. Phys.} {\bf 1966}, {\em 85}, 40.

\bibitem{jensen_1995}
Hjorth-Jensen, M.; Kuo, T.T.S.; Osnes, E. 
Realistic effective interactions for nuclear systems.
{\em Phys. Rep.} {\bf 1995}, {\em 261}, 125–270.

\bibitem{khint}
Brown, B.A.;
Double-octupole states in $^{208}$Pb.
{\em Phys. Rev. Lett.} {\bf 2000}, {\em 85}, 5300.

\bibitem{sn100pn}
Brown, B.A.; Stone, N.J.; Stone, J.R.; Towner, I.S.; Hjorth-Jensen, M.
Magnetic moments of the 2$^+_1$ states around $^{132}$Sn.
{\em Phys. Rev. C} {\bf 2005}, {\em 71}, 044317.

\bibitem{lnps}
Lenzi, S.M.; Nowacki, F.; Poves, A.; Sieja, K.
Island of inversion around $^{64}$Cr.
{\em Phys. Rev. C} {\bf 2010}, {\em 82}, 054301.

\bibitem{m3y}
Bertsch, G.; Borysowicz, J.; McManus, H.; Love, W.G.
Interactions for inelstic scattering derived from realistic potentials.
{\em Nucl. Phys. A}  {\bf 1977}, {\em 284}, 399-419.

\bibitem{tensor1}
Osterfeld, F.
Nuclear spin and isospin excitations.
{\em Rev. Mod. Phys.} {\bf 1992}, {\em 64}, 491-557.

\bibitem{tensor2}
B\"ackman, S.-O.; Brown, G.E.; Niskanen, J.A.,  
The nucleon-nucleon interaction and the nuclear many-body problem.
{\em Phys. Rep.} {\bf 1985}, {\em 124}, 1-68.

\bibitem{otsuka_2010b}
Otsuka, T., Suzuki, T., Holt, J. D., Schwenk, A., and Akaishi, Y., 
Three-body forces and the limit of oxygen isotopes.
{\em Phys. Rev. Lett.} {\bf 2010}, {\em 105}, 032501.

\bibitem{QMC_RMP}
Carlson, J.; {\it et al.}, 
Quantum Monte Carlo methods for nuclear physics.
{\em Rev. Mod. Phys.} {\bf 2015}, {\em 87}, 1067-1118.

\bibitem{dripline2020}
Tsunoda, N.; Otsuka, T.; Takayanagi, K.; Shimizu, N.; Suzuki, T.; Utsuno, Y.; Yoshida, S.; Ueno, H.
The impact of nuclear shape on the emergence of the neutron dripline.
{\em Nature} {\bf 2020}, {\em 587}, 66-71.

\bibitem{otsuka_shapecoexi} 
Otsuka, T.; Tsunoda, Y.
The role of shell evolution in shape coexistence.
{\em J. Phys. G} {\bf 2016}, {\em 43}, 024009.

\bibitem{otsuka_2001}
Otsuka, T.; Fujimoto, R.; Utsuno, Y.; Brown, B.A.; Honma, M.; Mizusaki, T.
Magic Numbers in Exotic Nuclei and Spin-Isospin Properties of the {\it NN} Interaction.
{\em Phys. Rev. Lett.} {\bf 2001}, {\em 87}, 082502.

\bibitem{janssens_2005}
Janssens, R.V.F.
Elusive magic numbers.
{\em Nature} {\bf 2005}, {\em 435}, 897-898.

\bibitem{steppenbeck_2013}
Steppenbeck, D., {\it et al.}
Evidence for a new nuclear `magic number' from the level structure of $^{54}$Ca.
{\em Nature} {\bf 2013}, {\em 502}, 207-210.

\bibitem{michimasa2018}
Michimasa, S.; {\it et al.}
Magic Nature of Neutrons in 54Ca: First Mass Measurements of $^{55–57}$Ca.
{\em Phys. Rev. Lett.} {\bf 2018}, {\em 121}, 022506.

\bibitem{chen2019}
S. Chen {\it et al.}, 
Quasifree Neutron Knockout from $^{54}$Ca  Corroborates Arising 
$N$=34 Neutron Magic Number.
{\em Phys. Rev. Lett.} {\bf 2019}, {\em 123}, 142501.

\bibitem{schiffer_2004}
J. P. Schiffer {\it et al.}, 
Is the Nuclear Spin-Orbit Interaction Changing with Neutron Excess?
{\em Phys. Rev. Lett.} {\bf 2004}, {\em 92}, 162501.

\bibitem{sahin_2017}
Sahin, E.; {\it et al.}
Shell Evolution towards $^{78}$Ni: Low-Lying States in $^{77}$Cu. 
{\em Phys. Rev. Lett.} {\bf 2017}, {\em 118}, 242502.

\bibitem{ichikawa_2019}
Ichikawa, Y.; {\it et al.}
Interplay between nuclear shell evolution and shape deformation revealed by the magnetic
moment of $^{75}$Cu,
{\em Nature Physics} {\bf 2019}, {\em 15}, 321-325.

\bibitem{liddick2006}
Liddick, S. N.; {\it et al.} 
Discovery of $^{109}$Xe and $^{105}$Te: Superallowed Decay near Doubly Magic $^{100}$Sn.
{\em Phys. Rev. Lett.} {\bf 2006}, {\em 97}, 082501.

\bibitem{seweryniak2007}
Seweryniak, D.;  {\it et al.}
Single-Neutron States in $^{101}$Sn.
{\em Phys. Rev. Lett.} {\bf 2007}, {\em 99}, 022504.

\bibitem{ensdf}
“Evaluated nuclear structure data file,” 
http://www.nndc.bnl.gov/ensdf/.

\bibitem{smirnova2010}
Smirnova, N.A.;  Bally, B.; Heyde, K.; Nowacki, F.; Sieja, K.  
Shell evolution and nuclear forces.
{\em Phys. Lett. B} {\bf 2010}, {\em 686}, 109-113.

\bibitem{kay2017}
Kay, B.P.; Hoffman, C.R.; Macchiavelli, A.O. 
Effect of Weak Binding on the Apparent Spin-Orbit Splitting in Nuclei.
{\em Phys. Rev. Lett.} {\bf 2017}, {\em 119}, 182502.

\bibitem{uozumi1994}
Uozumi, Y.; {\it et al.} 
Shell-model study of $^{40}$Ca with the 56-MeV ($\vec{d}, p$) reaction.
{\em Phys. Rev. C} {\bf 1994}, {\em 50}, 263-274.

\bibitem{burgunder2014}
Burgunder, G.; {\it et al.} 
Experimental Study of the Two-Body Spin-Orbit Force in Nuclei.
{\em Phys. Rev. Lett.} {\bf 2014}, {\em 112}, 042502.

\bibitem{tsunoda2014}
Tsunoda, Y.; Otsuka, T.; Shimizu, N.; Honma, M.; Utsuno, Y. 
Novel shape evolution in exotic Ni isotopes and configuration-dependent shell structure. 
{\em Phys. Rev. C} \textbf{2014}, {\em 89}, 031301(R).

\bibitem{mcsm1}
Honma, M.; Mizusaki, T.; Otsuka, T.
Diagonalization of Hamiltonians for Many-Body Systems by Auxiliary Field Quantum Monte Carlo Technique.
{\em Phys. Rev. Lett.} {\bf 1995}, {\em 75}, 1284.

\bibitem{mcsm2}
Otsuka, T., Mizusaki, T., and Honma, M.,
Structure of the $N$=$Z$=28 Closed Shell Studied by Monte Carlo Shell Model Calculation.
{\em Phys. Rev. Lett.} {\bf 1998}, {\em 81}, 1588-1591.

\bibitem{mcsm3}
Otsuka, T.; Honma, M.; Mizusaki, T.; Shimizu, N.; Utsuno, Y. 
Monte Carlo Shell Model for Atomic Nuclei.
{\em Prog. Part. Nucl. Phys.} {\bf 2001}, {\em 47}, 319-400.

\bibitem{mcsm4}
Shimizu, N.; Abe, T.; Tsunoda, Y.; Utsuno, Y.; Yoshida, T.; Mizusaki, T.; Honma,  M.; Otsuka, T.
New-generation Monte Carlo shell model for the K computer era.
{\em Prog. Theor. Exp. Phys.} \textbf{2012}, {\em 2012} 01A205.

\bibitem{otsuka2016}
Otsuka, T.; Tsunoda, Y.
The role of shell evolution in shape coexistence,
{\em J. Phys. G,} \textbf{2016}, {\em 43}, 024009.

\bibitem{leoni_2017}
Leoni, S.; {\it et al.}
Multifaceted Quadruplet of Low-Lying Spin-Zero States in $^{66}$Ni: Emergence of Shape Isomerism in Light Nuclei.
{\em Phys. Rev. Lett.} {\bf 2017},{\em 118}, 162502.

\bibitem{heyde2011} 
Heyde, K.; Wood, J. L.
Shape coexistence in atomic nuclei.
{\em Rev. Mod. Phys.} {\bf 2011}, {\em 83}, 1467-1521.

\bibitem{togashi_2016}
Togashi, T.; Tsunoda, Y.; Otsuka, T.; Shimizu, N.
Quantum Phase Transition in the Shape of Zr isotopes.
{\em Phys. Rev. Lett.} {\bf 2016}, {\bf 117}, 172502.

\bibitem{kremer_2016}
Kremer, C., {\it et al.},
First Measurement of Collectivity of Coexisting Shapes Based on Type II Shell Evolution: The Case of $^{96}$Zr.
{\em Phys. Rev. Lett.} {\bf 2016}, {\em 117}, 172503.

\bibitem{singh_2018}
Singh, P.; {\it et al.}
Evidence for Coexisting Shapes through Lifetime Measurements in $^{98}$Zr.
{\em Phys. Rev. Lett.} {\bf 2016},  {\em 121}, 192501.

\bibitem{togashi_2018}
Togashi, T.; Tsunoda, Y.; Otsuka, T.; Shimizu, N.; and Honma, M.
Novel Shape Evolution in Sn Isotopes from Magic Numbers 50 to 82.
{\em Phys. Rev. Lett.} {\bf 2018}, {\em 121}, 062501.
125
\bibitem{marsh_2018}
Marsh, B.A.; {\it et al.}
Characterization of the shape-staggering effect in mercury nuclei.
{\em Nature Physics} {\bf 2018}, {\em 14}, 1163-1167.

\bibitem{marginean_2020}
M\u{a}rginean, S., {\it et al.},
Shape Coexistence at Zero Spin in $^{64}$Ni Driven by the Monopole Tensor Interaction.
{\em Phys. Rev. Lett.} {\bf 2020},{\em 125}, 102502.

\bibitem{otsuka_2019}
Otsuka, T.; Tsunoda, Y.; Abe, T.; Shimizu, N.; Van Duppen, P.
Underlying Structure of Collective Bands and Self-Organization in Quantum Systems.
{\em Phys. Rev. Lett.} {\bf 2019}, {\em 123}, 222502.

\bibitem{robeldo_2008}
Robeldo, L.M.; {\it et al.}
Evolution of nuclear shapes in medium mass isotopes from a microscopic perspective.
{\em Phys. Rev. C} {\bf 2008},{\em 78}, 034314.

\bibitem{li_2009}
Li, ., {\it et al.},
Microscopic analysis of nuclear quantum phase transitions in the $N\approx$ 90 region.
{\em Phys. Rev. C} {\bf 2009},{\em 79}, 054301.

\bibitem{delaroche_2010}  
Delaroche, J.-P.; {\it et al.}
Structure of even-even nuclei using a mapped collective Hamiltonian and the D1S Gogny interaction
{\em Phys. Rev. C} {\bf 2010},{\em 81}, 104303.

\bibitem{tsunoda_2021}
Tsunoda, Y.; Otsuka, T.
Triaxial rigidity of $^{166}$Er and its Bohr-model realization.
{\em Phys. Rev. C} {\bf 2021},{\em 103}, L021303.

\bibitem{davydov1}
Davydov, A.S.; Filippov, G.F.  
Rotational states in even atomic nuclei.
{\em Nucl. Phys.} {\bf 1958},{\em 8}, 237-249. 

\bibitem{davydov2}
Davydov, A.S.; Rostovsky, V.S.
Relative transition probabilities between rotational levels of non-axial nuclei.
{\em Nucl. Phys.} {\bf 1959},{\em 12}, 58-68.  

\bibitem{sun2000}
Sun, Y.; {\it et al.}
Multiphonon $\gamma$-vibrational bands and the triaxial projected shell model.
{\em Phys. Rev. C} {\bf 2000},{\em 61}, 064323.

\bibitem{sun2002}
Boutachkov, P.; Aprahamian, A.; Sun, Y.; Sheikh, J.A.; Frauendorf, S. 
In-band and inter-band B(E2) values within the Triaxial Projected Shell Model.
{\em Eur. Phys. J. A} {\bf 2002}, {\em 15}, 455.

\bibitem{hayashi_1984}
Hayashi, A.; Hara, K.; Ring, P. 
Existence of Triaxial Shapes in Transitional Nuclei.
{\em Phys. Rev. Lett.} {\bf 1984}, {\em 53}, 337-340.

\bibitem{stone2016}
Stone, N. J.,  
Table of nuclear electric quadrupole moments.
{\em At. Data Nucl. Data Tables} {\bf 2016}, {\em 11-112}, 1-28.

\bibitem{nilsson1955}
Nilsson, S.G. 
Binding states of individual nucleons in strongly deformed nuclei.
{\em Mat. Fys. Medd. Dan. Vid. Selsk.} {\bf 1955},{\em 29}, no. 16.

\bibitem{kumar1968}
Kumar, K.; Baranger, M. 
Nuclear deformations in the pairing-plus-quadrupole model (III). Static nuclear shapes in the rare-earth region.
{\em Nucl. Phys. A} {\bf 1968},{\em 110}, 529-554.

\bibitem{bes1969}
Bes, D.R.; Sorensen, R.A. 
The Pairing-Plus-Quadrupole Model,
in {\em Advances in Nuclear Physics Vol. 2}; Baranger, M.; Vogt, E., Eds.; Plenum, New York, USA, 1969; Springer, Berlin, Germany 2012; Ch. 3.

\bibitem{elliott1958a}
Elliott, J.P.  
Collective motion in the nuclear shell model I.  Classification schemes for states of mixed configurations.  
{\em Proc. Roy. Soc.} {\bf 1958},{\em 245}, 128-145. 

\bibitem{elliott1958b}
Elliott, J.P.  
Collective motion in the nuclear shell model II.  The introduction of intrinsic wave-functions. 
{\em Proc. Roy. Soc.} {\bf 1958},{\em 245}, 562-581. 

\bibitem{caurier1994}
Caurier, E.; Zuker, A.; Poves, A.; Marti\'inez-Pinedo, G.
Full {\it pf} shell model study of {\it A}=48 nuclei.
{\em Phys. Rev. C} {\bf 1994}, {\em 50}, 225-236.

\bibitem{ahn2019}
Ahn, D.S.; {\it et al.} 
Location of the neutron dripline at fluorine and neon.
{\em Phys. Rev. Lett.} {\bf 2019}, {\em 123}, 212501.

\bibitem{n_tsunoda2017}
Tsunoda, N.; Otsuka, T.; Shimizu, N.; Hjorth-Jensen, M.; Takayanagi, K.; Suzuki, T.
Exotic neutron-rich medium-mass nuclei with realistic nuclear forces.
{\em Phys. Rev. C} {\bf 2017},{\em 95}, 021304(R).

\bibitem{machleidt2011}
Machleidt, R.; Entem, D.R. 
Chiral effective field theory and nuclear forces.
{\em Phys. Rep.} {\bf 2011}, {\em 503}, 1-75.

\bibitem{bogner_2002}
Bogner, S.; Kuo, T. T. S.; Coraggio, L.; Covello, A.; Itaco, N. 
Low momentum nucleon-nucleon potential and shell model effective interactions.
{\em Phys. Rev. C} {\bf 2002},{\em 65}, 051301.

\bibitem{nogga_2004}
Nogga, A.; Bogner, S.K.; Schwenk, A.
A. Low-momentum interaction in few-nucleon systems. 
{\em Phys. Rev. C} {\bf 2004},{\em 70}, 061002.

\bibitem{takayanagi2011a}
Takayanagi, K.
Effective interaction in non-degenerate model space. 
{\em Nucl. Phys. A} {\bf 2011},{\em 852}, 61–81.

\bibitem{takayanagi2011b}
Takayanagi, K.
Effective Hamiltonian in the extended Krenciglowa-Kuo method.
{\em Nucl. Phys. A} {\bf 2011},{\em 864}, 91–112.

\bibitem{n_tsunoda2014}
Tsunoda, N., Takayanagi, K., Hjorth-Jensen, M. and Otsuka, T., 
Multi-shell effective interactions.
{\em Phys. Rev. C} {\bf 2014},{\em 89}, 024313.

\bibitem{coraggio2009}
Coraggio, L.; Covello, A.; Gargano, A.; Itaco, N.
Similarity of nuclear structure in the $^{132}$Sn and $^{208}$Pb regions: Proton-neutron multiplets.
{\em Phys. Rev. C} {\bf 2009},{\em 80}, 021305.

\bibitem{arnswald2017}
Arnswald, K.; {\it et al.}
Enhanced collectivity along the $N$=$Z$ line: Lifetime measurements in $^{44}$Ti, $^{48}$Cr, and $^{52}$Fe.
{\em Phys. Lett. B} {\bf 2017},{\em 772}, 599–606.

\bibitem{tanihata1985}
Tanihata, I.; {\it et al.} 
Measurements of interaction cross sections and nuclear radii in the light p-shell region. 
{\em Phys. Rev. Lett.} {\bf 1985}, {\em 55}, 2676-2679.

\bibitem{hansen1987}
Hansen, P.G.; Jonson, B. 
The neutron halo of extremely neutron-rich nuclei. 
{\em Europhys. Lett.} {\bf 1987}, {\em 4}, 409-414.

\bibitem{abe2021}
Abe, T., {\it et al.}, 
Ground-state properties of light 4$n$ self-conjugate nuclei in {\it ab initio} no-core Monte Carlo shell model calculations with nonlocal {\it NN} interactions.
{\em Phys. Rev. C} {\bf 2021},{\em 104}, 054315.

\bibitem{otsuka2022}
Otsuka, T.; Abe, T.; Yoshida, T.; Tsunoda, Y.; Shimizu; N.; Itagaki, N.; Utsuno, Y.; Vary, J.; Maris, P.; Ueno, H.
{\em In preparation} {\bf 2022} {\em  }. 


\end{thebibliography}
\end{document}